\documentclass[11pt]{iopart}

\usepackage[utf8]{inputenc}

\usepackage{amsfonts, amssymb, latexsym}
\usepackage{graphicx}
\usepackage{rotating}
\usepackage[sort&compress,numbers]{natbib}
\usepackage{iopams}
\usepackage{xspace}

\eqnobysec

 \usepackage[usenames]{color}

\def\Lie{\mathcal{L}}
\def\A{\mathcal{X}}
\def\Aphi{\A_{\phi}}
\def\hAphi{\hat{\A}_{\phi}}
\def\E{\mathcal{E}}
\def\Ham{\mathcal{H}}
\def\M{\mathcal{M}}

\def\p{\partial}

\def\hg{\hat{\gamma}}
\def\hA{\hat{A}}
\def\hD{\hat{D}}
\def\hE{\hat{E}}
\def\hR{\hat{R}}

\def\hDelt{\hat{\triangle}}

\def\dif{{\rm{d}}}

\newcommand{\erf}{\textrm{erf}}

\usepackage[unicode,breaklinks]{hyperref}
\hypersetup{
    pdftitle={Nonlinear interactions between black holes and Proca fields},
    colorlinks=false,            
    bookmarks=true,              
  }

\begin{document}

\title[]{Nonlinear interactions between black holes and Proca fields}

\date{\today}

\author{
Miguel~Zilh\~ao$^{\dagger}$,
Helvi~Witek$^{\sharp, \flat}$,
Vitor~Cardoso$^{\star, \bullet}$ 
}

\address{$^{\dagger}$ Departament de F\'{\i}sica Fonamental \& Institut de Ci\`{e}ncies del Cosmos, Universitat de Barcelona, Mart\'{\i} i Franqu\`{e}s 1, E-08028 Barcelona, Spain}

\address{$^{\sharp}$ Department of Applied Mathematics and Theoretical Physics,
Centre for Mathematical Sciences, University of Cambridge,
Wilberforce Road, Cambridge CB3 0WA, UK}

\address{$^{\flat}$ School of Mathematical Sciences, University of Nottingham, University Park, Nottingham NG7 2RD, UK}

\address{$^{\star}$ CENTRA, Departamento de F\'{\i}sica, Instituto Superior T\'ecnico, Universidade de Lisboa, Avenida Rovisco Pais 1, 1049 Lisboa, Portugal}
\address{$^{\bullet}$ Perimeter Institute for Theoretical Physics Waterloo, Ontario N2J 2W9, Canada}

\eads{
\mailto{mzilhao@ffn.ub.es},
\mailto{h.witek@damtp.cam.ac.uk},
\mailto{vitor.cardoso@ist.utl.pt}
}

\begin{abstract}
Physics beyond the Standard Model is an important candidate for dark matter, and an interesting testing ground for strong-field gravity: the equivalence principle ``forces'' all forms of matter to fall in the same way, and it is therefore natural to look for imprints of these fields in regions with strong gravitational fields, such as compact stars or black holes. Here we study General Relativity minimally coupled to a massive vector field, and how black holes in this theory lose ``hair''. Our results indicate that black holes can sustain Proca field condensates for extremely long time-scales.
\end{abstract}


\section{Introduction}
\label{sec:intro}

Black holes (BHs) are strongly gravitating objects that solve the equations of General Relativity (GR)
and which are predicted to arise as the end-state of gravitational collapse. There is now overwhelming evidence
that the universe is populated by millions of BHs, in mass ranges of a few solar masses
to gigantic, billion solar-mass objects. Supermassive BHs are tightly connected to the evolution of their host galaxy
and may prompt or quench star formation in the entire galaxy through complex interactions between the BH and its environment. 

Black holes play a crucial role in fundamental physics, as the ``elementary particle'' of gravity.
This is largely due to the fact that both in GR and in many other modified theories of gravity, 
BHs are surprisingly simple objects characterized by only few parameters. 
In a series of rigidity, uniqueness and no-hair theorems~\cite{Bekenstein:1996pn,Carter:1997im,Chrusciel:2012jk}, 
it has been established that stationary BHs in Einstein-Maxwell theory
(in four spacetime dimensions) belong to the Kerr-Newman family of solutions and, thus, are completely characterized by their mass, angular momentum and (electric) charge.

Uniqueness properties together with the demonstration of mode stability kicked
off the success story of BHs as viable objects in the universe.  While
rigorous mathematical proofs of BH stability against generic perturbations are
still underway~\cite{Dafermos:2014cua}, it has been shown that subextremal BHs
are linearly stable against a wide class of perturbations involving
{\textit{massless}} fields~\cite{Whiting:1988vc,Berti:2009kk,Pani:2013wsa,
  Dias:2015wqa}.\footnote{Note, that extremal BH solutions are
  {\textit{unstable}}~\cite{Aretakis:2011ha}, but it is unlikely that these
  solutions are ever attained in realistic astrophysical
  environments~\cite{Thorne:1974ve}.}  These perturbative results are now being
complemented by studies concerning the nonlinear stability of
BHs~\cite{Zilhao:2014wqa,Cardoso:2014uka}, a property that is crucial for the
understanding of highly dynamical scenarios.

Black holes can be employed as sensitive ``scales'' to 
investigate ultra-light fundamental fields predicted in beyond-standard model
physics~\cite{Arvanitaki:2009fg,Kodama:2011zc,Jaeckel:2010ni}.
The latter has become possible with the realization that the paradigm ``BHs
in four-dimensional GR are stable'' is no longer true in the presence of
{\textit{massive}} bosonic fields.  Instead, BHs may suffer from a superradiant
or ``BH-bomb'' type instability~\cite{Press:1972zz,Brito:2015oca},
as has recently been proven rigourously~\cite{Shlapentokh-Rothman:2013ysa}~\footnote{At
  the onset of the instability a branch of new, hairy BH solutions has been
  found for complex scalars. These configurations evade standard no-hair
  theorems and essentially interpolate between Kerr BHs and boson
  stars~\cite{Herdeiro:2014goa,Herdeiro:2015gia}. The time-development of the
  superradiant instability in an adiabatic approximation, and the appearance of scalar condensates was
  investigated both for real and complex scalars in~\cite{Brito:2014wla}.}
This instability requires two ingredients:
(i)~Superradiance, the amplification
of low-frequency waves when they scatter off spinning BHs~\cite{Brito:2015oca}.
For monochromatic waves of frequency $\omega_R$ scattering off a spinning BH
with angular velocity $\Omega$ at the horizon, the condition is
\begin{equation}
  \label{eq:SRcondition}
  0 < \omega_{R} < m \Omega_{H}\,,
\end{equation}
where $m$ is the azimuthal ``quantum'' number of the wave.  This provides a
classical mechanism to amplify the impinging field at the expense of the BH's
rotational energy.
(ii)~Confinement of the field in the vicinity of the BH. As a consequence of
these two ingredients, the field is successively amplified leading to an
exponential growth of amplitude.

Massive bosonic fields with mass parameter $\mu=m_{B}/\hbar$ create such a confining cavity,
thus yielding superradiant instabilities~\cite{Damour:1976,Detweiler:1980uk,Zouros:1979iw}.
The timescale of the process is regulated by the coupling
\begin{equation}
\frac{G}{c\hbar} M\mu = 7.45\cdot 10^{9} \left(\frac{M}{M_{\odot}}\right) \left(\frac{m_{B}}{eV/c^2}\right)\,,
\end{equation}
between the BH mass $M$ and the field's mass $m_{B}$.
This coupling is too large for astrophysical BHs 
(with masses ranging between a few solar masses up to $10^{10}M_{\odot}$)
and standard model particles
-- consider, for example, the Higgs boson for which $M\mu\sim10^{21}\ldots10^{30}$ -- 
and yields instability timescales that are much longer than the age of the universe~\cite{Zouros:1979iw};
however, the ``BH-bomb'' mechanism becomes relevant
if $M\mu\sim\mathcal{O}(1)$ (in natural units).
Thus, astrophysical BHs are sensitive to
ultra-light bosonic fields in the mass range
$10^{-23} eV \lesssim m_{B}c^2 \lesssim 10^{-10} eV$.
%

These fields appear naturally when more fundamental theories 
that attempt to marry gravity with quantum physics
are compactified to four spacetime dimensions.
Depending on their interpretation, these fields may play a role as modifications of GR,
the simplest of which are scalar-tensor theories,
where the scalar field may acquire an effective mass due to interactions with the environment~\cite{Cardoso:2013opa,Berti:2015itd}.
Similarly, there are extensions of GR involving additional vector fields such as
Einstein-Aether theory~\cite{Jacobson:2000xp},
Horava-Lifshitz gravity~\cite{Horava:2009uw},
or higher dimensional gravity when compatified to four spacetime dimensions~\cite{Emparan:2008eg},
to name but a few explicit examples.
Although these additional vector fields are typically massless, environmental effects may induce an effective
mass term as described above for scalar-tensor theories.
Understanding the properties and phase-space of BHs in alternative models of gravity is an active field of research.
For example, no-hair theorems in generic scalar-tensor theories have been discussed in Sotirious' contribution
to this Focus Issue~\cite{Sotiriou:2015pka}.
Of direct interest for the applications in the present manuscript is Bekenstein's series of 
papers~\cite{Bekenstein:1971hc,Bekenstein:1972ky}
in which it was shown that static and stationary BHs cannot be bestowed with massive scalar, 
vector or spin-2 fields. This implies that, e.g.,  the Schwarzschild solution is the unique static solution of the 
Einstein-Proca system.
An exception are the recently constructed ``hairy BHs'' that appear at the onset of the instability and
evade these no-hair theorems~\cite{Herdeiro:2014goa,Herdeiro:2015gia}.
In the context of particle physics, the ``axiverse'' scenario
suggests the existence of an entire landscape of axion-like particles, i.e., ultra-light 
pseudo-scalar fields that become massive through spontaneous symmetry breaking 
or nonperturbative effects~\cite{Arvanitaki:2009fg,Kodama:2011zc,Jaeckel:2010ni}.
Another natural extension of the standard model involves additional vector fields.
In particular, compactifications of string theories 
(or their low-energy incarnations)
yield a hidden $U(1)$ gauge group whose symmetry may be broken 
via a Higgs-like mechanism~\cite{Jaeckel:2010ni,Goodsell:2009xc}.
This spontaneous symmetry breaking may endow the hidden ``photon''
with a mass,
in a manner similar to the standard-model Higgs mechanism that is responsible for the mass of the 
$W^{\pm}$ and $Z$ bosons.

In summary, superradiant instabilities can turn astrophysical BHs
into ``laboratories'' to investigate the properties of these fundamental
fields.
Potentially observable signatures include
(i) gaps in the Regge plane, i.e. the BHs' mass-spin phase-space~\cite{Arvanitaki:2009fg,Brito:2014wla};
(ii) modifications of the inspiral dynamics and characteristic, monochromatic gravitational wave (GW) 
emission~\cite{Okawa:2014nda,Arvanitaki:2014wva,Rosa:2015hoa,Alsing:2011er};
or (iii) gravitational radiation induced by a bosenova collapse~\cite{Yoshino:2012kn}.

With this motivation in mind, the phenomenology of
massive fields in BH backgrounds has been an active field of research~\cite{Brito:2015oca}.
%
Unfortunately, the timescales involved are so large that most studies have focused on perturbative calculations
of the instability. Recently, we started a long-term programme aimed at understanding the nonlinear development
of fundamental fields around spinning BHs, the first step of which focused on massive scalars~\cite{Okawa:2014nda}.
A natural next step---and our purpose herein---is the extension of these studies to massive vector, i.e. Proca fields.
Frequency-domain, linearized calculations in the background of Schwarzschild BHs~\cite{Rosa:2011my,Galtsov:1984nb}
and slowly-rotating BH spacetimes~\cite{Pani:2012bp,Pani:2012vp}
revealed a hydrogen-like frequency spectrum in the small-coupling limit.
These results have been used to put new stringent bounds 
on the allowed mass of the hidden ``photon'', $\mu_{\gamma}\lesssim10^{-20}eV$~\cite{Pani:2012vp}.
The growth rates for the superradiant instability of massive vector fields
are up to four orders of magnitudes larger than those of the massive scalar~\cite{Pani:2012bp}
and linearized, time-domain studies have predicted timescales of
$\tau = 3\cdot10^{3} \frac{G M}{c^3} \sim 0.01 s \frac{M}{M_{\odot}}$
in the best case scenario~\cite{Witek:2012tr}.
For this reason Proca fields represent an appealing model to explore the ``BH-bomb'' instability 
at the fully nonlinear level.

The present paper is the first installment of a series attending this challenging issue.
Here, we focus on developing the formalism and presenting a full-blown, $3+1$-dimensional numerical
code capable of evolving BH spacetimes coupled to massive vector fields.
Our results have been obtained through two different, independent implementations.
For the sake of testing and benchmarking our codes we restrict ourselves to the evolution of massive vector
fields in a nonrotating spacetime. The evolution of spinning geometries will be considered elsewhere.

The paper is organized as follows:
In section~\ref{sec:model} we introduce our model
and present its time evolution formulation.
We derive constraint-preserving initial configurations in section~\ref{sec:init-data}.
In section~\ref{sec:WaveExtraction} we provide our formalism to extract information about the emitted radiation.
In section~\ref{sec:results} we give a brief summary of our implementation and present the numerical results. 
We finish with our conclusions in section~\ref{sec:final}.
%
%
Throughout the paper we use natural units, i.e., $c=1=G$ and $\hbar=1$. 
Greek letters $\mu,\nu, \ldots$ refer to spacetime indices running from $0,\ldots,3$,
whereas Latin letters $i,j,\ldots$ denote spatial indices running from $1,\ldots,3$.
%

\section{The model -- coupling Proca fields to gravity}
\label{sec:model}
\subsection{Action and equations of motion}
\label{ssec:eom}

We consider an action coupling a vector field $X_{\mu}$ with mass 
$m_{\rm{V}} = \mu_{\rm{V}} \hbar$ to gravity~\cite{Arvanitaki:2009fg,Jaeckel:2010ni}
\begin{equation}
\label{eq:Action}
S = \int d^4x \sqrt{-g} 
      \left( \frac{R}{4} - \frac{1}{4}W^{\mu\nu}W_{\mu\nu} - \frac{\mu_{\rm V}^2}{2}X_{\nu}X^{\nu} \right)
\,,
\end{equation}
where 
$W_{\mu\nu}=\nabla_{\mu}X_{\nu} - \nabla_{\nu}X_{\mu}$.
Note, that we consider solely the gravitational coupling to massive vector fields,
because we ultimately aim at exploring superradiant scattering, a purely gravitational effect
that is independent of the kinematic mixing between the hidden sector and visible photons.
The resulting equations of motion (EoMs) are
\begin{eqnarray}
\label{eq:EoMVector}
&&\nabla_{\nu} W^{\mu\nu} + \mu^{2}_{\rm V} X^{\mu}=0\,, \\
\label{eq:EoMGR}
&&R_{\mu\nu} - \frac{1}{2}g_{\mu\nu} R = 2 W_{\mu\rho} W_{\nu}{}^{\rho} - \frac{1}{2} g_{\mu\nu} W^{\rho\sigma} W_{\rho\sigma}
       + \mu^{2}_{\rm V} \left( 2 X_{\mu} X_{\nu} - g_{\mu\nu} X^{\rho} X_{\rho} \right) \,.
\end{eqnarray}
In contrast to Einstein-Maxwell theory (where $\mu_{\rm V} = 0$), 
the vector $X_{\mu}$ is not a pure gauge field
but plays the role of a fundamental, physically meaningful field.
This implies that the Lorenz condition 
\begin{equation}
\label{eq:LGmod}
\nabla^{\mu} X_{\mu} = 0
\,,
\end{equation} 
has to be satisfied, as can be seen from~\eref{eq:EoMVector}.

\subsection{Formulation as a Cauchy problem}
%
To explore the nonlinear interactions between Proca fields and BHs 
we have to track their dynamics numerically.
Therefore, we recast the EoMs~\eref{eq:EoMVector} and~\eref{eq:EoMGR} as a time evolution problem.
We begin by foliating the spacetime into $3$-dimensional spatial slices $\Sigma_{t}$ with
metric $\gamma_{ij}$. 
The $3$-metric defines the projection operator 
\begin{equation}
\label{eq:ProjOp}
\gamma^{\mu}{}_{\nu} = \delta^{\mu}{}_{\nu} + n^{\mu} n_{\nu}
\,,
\end{equation}
where $n^{\mu}$ denotes the timelike unit vector (normal to $\Sigma_{t}$)
with normalization $n^{\mu} n_{\mu}=-1$.
The full, $4$-dimensional spacetime metric $g_{\mu\nu}$ can then be expressed as
\begin{equation}
\label{eq:LineElement}
\dif s^{2} = g_{\mu\nu} \dif x^{\mu} \dif x^{\nu}
           = - \left( \alpha^{2} - \beta^{i} \beta_{i} \right) \dif t^{2}
             + 2 \gamma_{ij} \beta^{i} \dif t \dif x^{j}
             +   \gamma_{ij} \dif x^{i} \dif x^{j}\,, 
\end{equation}
where the lapse function $\alpha$ and shift vector $\beta^{i}$ describe the coordinate degrees of freedom.
To complete the characterization of the full spacetime we define the extrinsic curvature
\begin{equation}
\label{eq:KijDef}
K_{ij}  =   - \frac{1}{2\alpha} \left( \partial_{t} - \Lie_{\beta} \right) \gamma_{ij} 
\,.
\end{equation}
%
We now split the Proca field into its scalar $\Aphi$ and $3$-vector $\A_{i}$ potential given by
\begin{eqnarray}
\label{eq:VecSplit}
X_{\mu} =  \A_{\mu} + n_{\mu} \Aphi
\,,\quad{\rm{with}}\quad
\A_{i} = \gamma^{\mu}{}_{i} X_{\mu}
\,,\quad{\rm{and}}\quad
\Aphi = - n^{\mu} X_{\mu}
\,.
\end{eqnarray}
In analogy to Maxwell's theory we introduce an ``electric'' and ``magnetic'' field,
%
\begin{equation}
\label{eq:DefEB}
E_{i} = \gamma^{\mu}{}_{i} W_{\mu\nu} n^{\nu}
\,,\quad
B_{i} = \gamma^{\mu}{}_{i} \,^{\ast} W_{\mu\nu} n^{\nu} = \epsilon^{ijk} D_{j} \A_{k}
\end{equation}
where $E_{\mu} n^{\mu} = 0$ and $B_{\mu} n^{\mu} = 0$ hold by definition.
As we will see later, the former definitions provide a time evolution equation for the $3$-vector potential,
while we employ the latter purely as {\textit{short-hand notation}}.
%
The $W_{\mu \nu}$ tensor is reconstructed from
\begin{equation}
\label{eq:FmninEA}
W_{\mu\nu}= n_{\mu} E_{\nu} - n_{\nu} E_{\mu} + D_{\mu} \A_{\nu} - D_{\nu} \A_{\mu}\,, 
\end{equation}
where $D_{\mu}$ is the covariant derivative with respect to the $3$-metric.

The definition of the extrinsic curvature and the $W_{\mu \nu}$ tensor prescribe the time evolution of the 
$3$-metric $\gamma_{ij}$ and (spatial) vector potential $\A_{i}$, whereas the dynamics of the model are determined by the EoMs~\eref{eq:EoMVector} and~\eref{eq:EoMGR}. Their spatial projection
yields the evolution equation for the ``electric'' field and extrinsic curvature.
Additionally, the Lorenz condition~\eref{eq:LGmod} prescribes the evolution of 
the scalar potential $\Aphi$.

Proca's equations (just like its Maxwellian counterparts) give rise to constraints which the evolution equations are subjected to---the familiar ``Gauss'' constraint. 
However, analysing the evolution partial differential equations (PDEs) reveals the existence of a mode with zero characteristic speed.
In other words, a constraint violation (which is always present in numerical simulations)
will not be propagated away, but will rather stay in the numerical domain.
It is therefore desirable to modify the Proca evolution equations such that all characteristic speeds are 
non-zero and the constraints are damped\footnote{
A similar procedure for the GR evolution sector also exists and is known as the ``Z4'' formulation~\cite{Bona:2003qn,Bernuzzi:2009ex,Alic:2011gg}.
}.
Guided by numerical simulations of BH binaries in Einstein-Maxwell theory 
we have introduced an additional variable $Z$ and damping parameter $\kappa>0$,
which will help stabilising the numerical evolution~\cite{Hilditch:2013sba, Gundlach:2005eh,Palenzuela:2009hx}. 
We recover the original system~\eref{eq:EoMVector},~\eref{eq:EoMGR} and~\eref{eq:LGmod}
by setting $Z=0$.
In summary, the GR-Proca model evolves according to
\begin{eqnarray}
\label{eq:dtgamma}
\p_{t} \gamma_{ij} & = & - 2 \alpha K_{ij} + \Lie_{\beta} \gamma_{ij}
,\\
\label{eq:dtAi}
\p_{t} \A_{i}      & = & - \alpha \left( E_{i} + D_{i} \Aphi \right) - \Aphi D_{i}\alpha + \Lie_{\beta} \A_{i}
,\\
\label{eq:dtE}
\p_{t} E^{i}       & = &
        \alpha \left( K E^{i} + D^{i} Z + \mu^2_{\rm V} \A^{i}
                + \epsilon^{ijk} D_{j} B_{k} \right)
        - \epsilon^{ijk} B_{j} D_{k}\alpha
        + \Lie_{\beta} E^{i},\\
\label{eq:dtKij}
\p_{t} K_{ij}      & = & - D_{i} D_{j} \alpha
        + \alpha \left( R_{ij} - 2 K_{ik} K^{k}{}_{j} + K K_{ij} \right)
        + \Lie_{\beta} K_{ij} 
\nonumber \\ & &
        + 2 \alpha \left( E_{i} E_{j} - \frac{1}{2} \gamma_{ij} E^{k} E_{k} 
        + B_{i} B_{j} - \frac{1}{2} \gamma_{ij} B^{k} B_{k}
        - \mu^{2}_{\rm V} \A_{i} \A_{j}
          \right),\\
\label{eq:dtAphi}
\p_{t} \Aphi  & = & - \A^{i} D_{i} \alpha
        + \alpha \left( K \Aphi - D_{i} \A^{i} - Z \right)
        + \Lie_{\beta} \Aphi ,\\
\label{eq:dtZ}
\p_{t} Z          & = & \alpha \left( D_{i} E^{i} + \mu^{2}_{\rm V} \Aphi - \kappa Z \right)
        + \Lie_{\beta} Z\,,
\end{eqnarray}
where $\Lie$ denotes the Lie derivative.
Additionally, the evolution is subject to a set of constraints given by
\begin{eqnarray}
\label{eq:Hamiltonian}
\Ham & = R - K_{ij} K^{ij} + K^2 
        - 2 \left( E^{i} E_{i} + B^{i} B_{i} + \mu^{2}_{\rm V} \left( \Aphi^{2} + \A^{i} \A_{i} \right)
            \right)
       = 0\,,\\
\label{eq:momentumConstraint}
\M_{i} & = D^{j} K_{ij} - D_{i} K 
        - 2 \left( \epsilon_{ijk} E^{j} B^{k} + \mu^{2}_{\rm V} \Aphi \A_{i}
            \right)
       = 0
\,.
\end{eqnarray}
In practice, we adopt a free evolution scheme, in which the constraints~\eref{eq:Hamiltonian}, \eref{eq:momentumConstraint} are only solved for the initial data which will then be evolved.
When setting up the relevant initial data, we also need to satisfy the ``Gauss'' constraint
\begin{eqnarray}
\label{eq:DivEConstraint}
\E   = D_{i} E^{i} + \mu^{2}_{\rm V} \Aphi     = 0 \,,
\end{eqnarray}
which is enforced during the evolution through~\eref{eq:dtZ}.

Throughout the numerical simulation we will not actually use the ADM-York type formulation, given by~\eref{eq:dtgamma}--\eref{eq:dtZ}, which is an ill-posed Cauchy problem, but we will employ the
Baumgarte-Shapiro-Shibata-Nakamura (BSSN) scheme~\cite{Shibata:1995we,Baumgarte:1998te} 
together with the moving puncture gauge~\cite{Brandt:1997tf,Campanelli:2005dd,Baker:2005vv}.
This reformulation involves a conformal decomposition
together with a constraint addition rendering it into a well-posed initial value problem~\cite{Alcubierre:2008,Witek:2010es}.
The explicit expressions are rather lengthy and therefore given in~\ref{app:BSSNeqs}.

\subsection{Flat space limit and Yukawa suppression}
\label{ssec:flat}

Minkowski geometry is a good approximation at large distance, and provides 
a useful benchmark for our results, in particular for the time and spatial decay 
of the fields. The \emph{ansatz}
\begin{equation}
X_\mu=\left(\int Q_1 dr,\int Q_2 dt,0,0\right)\,,
\end{equation}
where $Q_{i}=Q_{i}(t,r)$, 
produces the equation of motion for $\Psi\equiv Q_1-Q_2$,
\begin{equation}
\Psi''-\ddot{\Psi}+\frac{2}{r}\Psi'-\frac{(\mu_{\rm V}^2r^2+2)}{r^2}\Psi=0\,,
\end{equation}
where $\dot{}$ and $'$ denote derivatives with respect to $t$ and $r$.
We find the harmonic solutions 
\begin{equation}
\Psi = e^{-\imath \omega t} \left( C_1 J_{-2}\left(\imath r \sqrt{\mu_{\rm V}^2-\omega^2}\right)
        + C_2 Y_{-2}\left(\imath r \sqrt{\mu_{\rm V}^2 - \omega^2}\right) \right)
\,,
\end{equation} 
where $Y,J$ are Bessel functions of the first kind and $C_{i}$ are constants.

Another flat space solution (in $3+1$ form) is given by
\begin{eqnarray}
E^r(r)&=C e^{-kr} \left( \frac{1}{r^2} + \frac{k}{r}  \right) \cos \omega t\\
\A^r(r)&=-C \frac{ \omega}{\mu_{\rm V}^2} e^{-kr} \left( \frac{1}{r^2} + \frac{k}{r}  \right) \sin \omega t\\
\Aphi(r)&=C \frac{k^2}{\mu_{\rm V}^2} \frac{e^{-kr}}{r} \cos \omega t\,,
\end{eqnarray}
where $\omega = \sqrt{\mu_{\rm V}^2 - k^2}$.

Both solutions teach us that at large distances the fields have a Yukawa-like
dependence on the radial distance for low-energy modes ($\omega<\mu_{\rm V}$)
and an oscillatory behaviour for high-energy modes.  Guided by the flat-space
solutions we expect to recover a Yukawa-like dependence for quasi-bound states
around BHs, because by definition these modes are trapped in the vicinity of the
BH.

\subsection{Summary of perturbative results}
\label{ssec:LiteratureSummary}

The above results concerning Yukawa suppression in Minkowski backgrounds are in fact shown to arise
also in a perturbative framework of (low-amplitude) Proca fields around BHs.
These results will be important as benchmarks for our numerical implementation of the nonlinear BH-Proca system.
For this purpose, we briefly summarize the most important results on the subject.

In contrast to Einstein-Maxwell theory, massive spin-1 fields propagate three physical degree of freedom.
These come in the shape of an axial mode (labelled with $S=0$)
and two polar modes with spin $S=\pm1$. 
Note, that the Proca field also propagates a monopole, i.e., spherically symmetric mode corresponding to $S=+1$.

In a perturbative approach, the field equations are linearized in a fixed-BH background and any function 
(e.g., the vector potential) can then be Fourier-decomposed $\sim e^{-\imath \omega t}$. 
When boundary conditions are imposed, the field equations become an eigenvalue problem for the characteristic frequencies, 
which become complex quantities, $\omega=\omega_R+\imath \omega_I$.
Frequency domain analysis of the Proca equation in the background of Schwarzschild BHs~\cite{Galtsov:1984nb,Rosa:2011my,Herdeiro:2011uu,Wang:2012tk,Sampaio:2014swa}
and slowly rotating BH spacetimes~\cite{Pani:2012vp,Pani:2012bp}
have revealed two classes of solutions that depend on the particular choice of the boundary conditions at spatial infinity.
One type of solution corresponds to quasi-normal modes (QNMs)~\cite{Berti:2009kk}, while the second class
are quasi-bound states (QBSs), which are particular of massive fluctuations and, essentially, decay exponentially at large distances.
Quasi-bound states require confined states, as we discussed previously, and only occur for $\omega_R \lesssim \mu_{\rm V}$.


For small mass couplings $M\mu_{\rm V}\ll 1$, the QBS are well described by
hydrogen-type spectra~\cite{Brito:2015oca},
\begin{equation}
\omega_{R} \sim \mu_{\rm V} \left( 1 - \frac{(M\mu_{\rm V})^{2} }{ 2 (l+n+S+1)^{2} } \right)\,,\label{hydrogen_spectra_vector}
\end{equation}
where 
$n\geq0$ is an overtone number and $l$ is the angular quantum number~\cite{Galtsov:1984nb,Rosa:2011my,Pani:2012bp}.
The growth or decay rate of the QBSs is encoded in the imaginary part of their frequencies
approximated by the analytic expression
\begin{equation}
\omega_{I} \sim (m a/M - 2 r_{+} \mu_{\rm V}) \left( M \mu_{\rm V} \right)^{4 l + 5 + 2S }
\,,
\end{equation}
in the slow-rotation regime~\cite{Pani:2012bp}. 
In the limit $a/M\rightarrow0$ it reduces to the results found in reference~\cite{Rosa:2011my} for 
the Schwarzschild case.
The superradiant instability is strongest in nearly extremal backgrounds and 
numerical computations in the time domain found a maximum growth rate of 
$M \omega_{I} \sim 5 \cdot 10^{-4}$
(for dimensionless spin parameter $a/M=0.99$)
which corresponds to an e-folding timescale of $\tau\sim0.01s \frac{M}{M_{\odot}}$~\cite{Witek:2012tr}.
%

The backscattering off the spacetime curvature gives rise to oscillatory power-law tails 
whose frequency is determined by the mass of the field
\begin{equation}
\label{eq:MassiveTails}
\psi \sim t^{p} \sin(\mu_{\rm V} t) \,,
\end{equation}
with the exponent $p=-(l+3/2+S) $ at intermediate late times $M\mu_{\rm V} \ll t\mu \ll \frac{1}{(M\mu_{\rm V})^2}$
and a universal decay $p=-5/6$ at very late times $\frac{1}{(M\mu_{\rm V})^2} \ll t\mu_{\rm V}$~\cite{Konoplya:2006gq,
Koyama:2001qw,Witek:2012tr}.
Thus, for monopole and dipole fluctuations the field is expected to decay, respectively, as 
$t^{-5/2} \sin(\mu_{\rm V} t)$ and $t^{-(5/2+S)} \sin(\mu_{\rm V} t) $ at intermediate times 
and to follow the universal behaviour $t^{-5/6} \sin(\mu_{\rm V} t)$ at very late-times, 
if the ``tail-stage'' is not superposed by the exponential, but slowly decaying bound-state stage.
Generic initial data will excite all possible modes. Whether tails or QBS dominate the signal depends on the type of initial data and
at which stage the signal is being observed.

\section{Initial data}
\label{sec:init-data}

In this section we describe the construction of initial configurations representing a BH
in the presence of a massive vector field.
Although we only evolve nonrotating BH spacetimes in this paper, our construction in principle allows for rotating solutions.
We intend to explore fields in the {\textit{nonperturbative}} regime. 
Then, an important ingredient for long-term stable
nonlinear simulations is the construction of {\textit{constraint-satisfying}} initial data~\cite{Okawa:2014nda}.
For the purpose of solving the initial data problem we employ the 
York-Lichnerowicz conformal decomposition~\cite{Lichnerowicz1944,York:1971hw}
\begin{eqnarray}
\fl \gamma_{ij} & = &\psi^4 \hat \gamma_{ij} 
\,, \quad
K^{ij}        =  \psi^{-10} \hat A^{ij} + \frac{1}{3} \psi^{-4} \hat \gamma^{ij} K
\,, \quad
E^i          = \psi^{-6} \hat E^i 
\,, \quad
\Aphi      = \psi^{-6} \hAphi
\,.
\end{eqnarray}
The constraints become 
\begin{eqnarray}
\label{eq:HamiltonianIDgen}
\Ham & = & \hDelt\psi - \frac{\psi}{8} \hR + \frac{1}{8\psi^7} \hA^{ij} \hA_{ij} - \frac{\psi^{5}}{12} K^2 
        + \frac{\mu^{2}_{\rm V}}{4\psi^{7}} \left( \hAphi^{2} + \psi^{8} \hg^{ij} \A_{i} \A_{j} \right)
\nonumber\\ & &
        + \frac{1}{4\psi^3} \hE_{i} \hE^{i}
        + \frac{1}{4\psi^3} \hD^{j} \A^{i} \left( \hD_{j} \A_{i} - \hD_{i} \A_{j} \right)
\,,\\
\label{eq:MomentumIDgen}
\M_{i} & = &     \hD^{j} \hA_{ij} - \frac{2}{3} \psi^{6} \hD_{i} K 
           + 2 \hE^{j} \left( \hD_{j} \A_{i} - \hD_{i} \A_{j} \right)
           - 2 \mu^{2}_{\rm V} \hAphi \A_{i}
\,,\\
\label{eq:DivEIDgen}
\E     & = & \hD_{i} \hE^{i} + \mu^{2}_{\rm V} \hAphi 
\,,
\end{eqnarray}
where the hatted operators $\hDelt$ and $\hD_{i}$ are constructed from the conformal metric $\hat \gamma_{ij}$.
These PDEs are in general difficult to solve, so
we now simplify the constraint equations 
using the following assumptions
\begin{eqnarray}
\label{eq:AnsatzVarsGaussE}
\hg_{ij} = \delta_{ij}
\,,\quad
K = 0
\,,\quad
\A_{i}   =   0
\,,\quad
\hE^{i}  =  - \delta^{ij} \p_{j} V
\,.
\end{eqnarray}
%
yielding
\begin{eqnarray}
\label{MomIDGauss}
0 & = & \hD_j \hA^j{}_i 
\,, \\
\label{eq:HamiltonianIDGauss}
0 & = & \hDelt\psi + \frac{1}{8\psi^7} \hA^{ij} \hA_{ij} 
+ \frac{1}{4\psi^3} \delta^{ij} \p_i V \p_j V
        + \frac{\mu^{2}_{\rm V}}{4\psi^{7}} \hAphi^{2}
\,,\\
\label{eq:DivEIDGauss}
0 & = & \hDelt V - \mu^{2}_{\rm V} \hAphi
\,.
\end{eqnarray}
In the following we will outline two different procedures to solve these equations.

\subsection{Simple Initial Data solutions}
\label{ssec:ID-analytic}

A straight-forward way to solve the constraints~\eref{MomIDGauss}--\eref{eq:DivEIDGauss}
is provided by further imposing $\hAphi = 0$.
Equation~\eref{MomIDGauss} has the known Bowen-York solution~\cite{Bowen:1980yu}, 
whereas the Laplace equation $\hat \triangle V = 0$ is solved by a potential $V\sim \frac{1}{r}$. 
With these solutions the constraint equations are now decoupled, and one can solve~\eref{eq:HamiltonianIDGauss}
numerically for the conformal factor $\psi$ using standard methods.

If we additionally assume time-symmetric data, i.e.\ $\hA_{ij} = 0$,
the momentum constraint~\eref{MomIDGauss} is satisfied trivially
and we find exact, analytic solutions for equations~\eref{eq:DivEIDGauss} and~\eref{eq:HamiltonianIDGauss} (remember we are imposing $\hAphi = 0$)
given by~\cite{Brill:1963yv,Alcubierre:2009ij}
\begin{equation}
\label{eq:psi-exact}
\psi^2  = \left( 1 + \frac{M}{2r} \right)^2 - \frac{C^2}{4r^2} 
\,, \qquad
V = \frac{C}{r} 
\,,
\end{equation}
where $C$ is a constant. We term this a ``Simple Initial Data'' (SID) solution.
Using our \emph{ansatz}~\eref{eq:AnsatzVarsGaussE} we can read off the initial profile for the ``electric'' field,
\begin{equation}
\label{eq:Er-analytic}
E^i = \psi^{-6} C\frac{x^i}{r^3} \,,
\end{equation}
where $x^i=(x,y,z)$. This construction is fully analytic, thus very simple to implement and run. 
In the Einstein-Maxwell case, $\mu_{\rm V} = 0$, 
this reduces to an exact (static) solution describing a single Reissner-Nordstr\"om BH with charge $C$.
In the present case, however, $C$ has to be interpreted as the amplitude of the Proca field 
instead of an electromagnetic charge.
The major drawback of this construction is its restriction to spherically symmetric vector fields
``attached'' directly to the BH.
We now show an alternative construction that allows for more generic field configurations.

\subsection{Condensates around black holes: Gaussian Initial Data solutions}
\label{ssec:ID-Aphi}

To access physically more interesting cases, such as non-spherically symmetric field profiles
or ``clouds'' surrounding the BH in some distance,
we construct initial data representing a (vector field) condensate around a BH, for Gaussian-like initial data (GID).
After imposing~\eref{eq:AnsatzVarsGaussE} we still have the freedom to specify $\hAphi$ 
and, as before,
the momentum constraint~\eref{MomIDGauss} has the known Bowen-York solution~\cite{Bowen:1980yu}.
Our strategy now is as follows:
\begin{enumerate}
\item 
    first we solve the ``Gauss'' constraint~\eref{eq:DivEIDGauss}. 
    We specify the ``shape'' of the condensate as 
    \begin{eqnarray}
    \label{eq:AphiIDGaussGen}
    \hAphi        = C \frac{R(r)}{r^{p}} \Sigma(\theta,\varphi),
  \end{eqnarray}
  where
    \begin{eqnarray*}
    R(r)              = \exp\left(-\frac{(r-r_{0})^2}{\sigma^2} \right),\qquad
    \Sigma(\theta,\varphi) = \sum_{lm} c_{lm} Y_{lm}(\theta,\varphi)
    \end{eqnarray*}
    composed of a Gaussian wavepacket $R(r)$ (of width $\sigma$ centered around $r_{0}$)
    with amplitude $C$.
    The angular distribution is given by a superposition of spherical harmonics $Y_{lm}(\theta,\varphi)$
    such that $\hAphi$ is real.
    With this \emph{ansatz} we can solve the resulting Poisson equation analytically for $V$ and,
    hence, for the conformal field~$\hE^{i}$.
\item Now we insert the \emph{ansatz} for $\hAphi$ and the corresponding $V$ into the Hamiltonian constraint.
     Since we want to consider BH solutions
     we use the puncture approach~\cite{Brandt:1997tf} and set
     \begin{eqnarray}
     \label{eq:u-def}
     \psi = \psi_{\rm BL} + u
     \,,\qquad
     \psi_{\rm BL} = 1 +  \frac{M}{2 r}
     \,,
     \end{eqnarray}
     where $\psi_{\rm BL}$ denotes the Brill-Lindquist conformal factor for a BH
     of mass $M$ located at the origin and the correction $u$ is a smooth
     function. The Hamiltonian constraint~\eref{eq:HamiltonianIDGauss} becomes
     an elliptic equation for $u$ which we solve numerically by modifying
     the \textsc{TwoPunctures} code~\cite{Ansorg:2004ds}, which is part of the
     {\textsc{Einstein Toolkit}}~\cite{Loffler:2011ay,EinsteinToolkit:web}. Having been
     originally developed to calculate four-dimensional vacuum puncture data for
     binary BH configuration, \textsc{TwoPunctures} has proven to work equally
     well when modified to handle different
     scenarios~\cite{Zilhao:2011yc,Zilhao:2013nda,Ruchlin:2014zva,Okawa:2014nda}. 
     As we will see, this is also the case in our current configuration.
\end{enumerate}
We have to ensure that the source terms entering the Hamiltonian remain regular at
the puncture and, furthermore, that the resulting elliptic PDE is well-posed.
To convince ourselves of the latter property we linearize the equation around a known solution $\psi_0$, $\psi=\psi_{0} + \epsilon$, with $|\epsilon| \ll \psi_0$.
Then, the linearized Hamiltonian is
\begin{eqnarray}
\Ham_{\rm lin} = &  \hDelt \epsilon
                - \epsilon \left( 
                    \frac{7}{8\psi_0^8} \hA^{ij} \hA_{ij} 
                  + \frac{3}{4\psi^{4}_{0}} \hE^{i} \hE_{i} 
                  + \frac{7\mu^{2}_{\rm V}}{4\psi^{8}_{0}} \hAphi^{2} \right)
             =    \hDelt \epsilon - h \epsilon
             = 0
\,.
\end{eqnarray}
The existence of a unique solution requires $h>0$~\cite{ChoquetBruhat:1999yh,Gourgoulhon:2007ue}
which is clearly satisfied by our construction.

To further simplify this procedure we start by considering time-symmetric
configurations, i.e. $\hA_{ij} = 0$.  Note, that we are in no way limited to
such configurations; we merely choose to focus first on these simpler cases
before advancing to more involved examples.  We will now explicitly write
specific constructions that we have evolved.
\begin{description}
\item[Gaussian initial data I -- spherical symmetry:] 
We first focus on spherically symmetric initial data and set 
$\Sigma(\theta,\varphi)=2\sqrt{\pi} Y_{00} = 1$ and $p=1$ in~\eref{eq:AphiIDGaussGen}.
With this \emph{ansatz} we obtain
\begin{eqnarray}
\label{eq:SolGaussIDSphSym}
\fl
\hAphi = \frac{C_{00}}{r} R(r)
\,,\\
\fl
\hE^{r}   = - \frac{C_{00} \mu^{2}_{\rm V} \sigma^{2} }{2 r^2}
        \left[ \exp\left( - \frac{r^2_{0}}{\sigma^2} \right) - R(r)
             + \frac{\sqrt{\pi} r_{0} }{\sigma} 
               \left\{ \erf\left( \frac{r_{0}}{\sigma} \right) + \erf\left( \frac{r-r_{0}}{\sigma} \right)
               \right\}
        \right],\\
\fl
\hE^{\theta} =  0 
\,,\qquad
\hE^{\varphi} = 0 
\,,
\end{eqnarray}
where $R(r)$ is given in~\eref{eq:AphiIDGaussGen}.
We obtain the ``electric'' field in Cartesian coordinates by a standard coordinate transformation.
If we insert this solution into the source terms in~\eref{eq:HamiltonianIDGauss} and 
recall that $\psi\sim1/r$ we can convince ourselves that they remain regular at the puncture.

\item[Gaussian initial data II -- the dipole:]
We next construct initial data sourced by an axisymmetric dipole field, i.e., we set the angular function
$\Sigma(\theta,\varphi)=Y_{10}(\theta,\varphi)$ in~\eref{eq:AphiIDGaussGen} and choose the exponent $p=0$.
Specifically, we take the \emph{ansatz}
\begin{eqnarray}
\label{eq:GaussIDl1m0Aphi}
\hAphi = & C_{10} R(r) Y_{10}(\theta,\varphi) 
           =   C_{10} \sqrt{\frac{3}{4\pi}} R(r) \cos\theta 
\,,
\end{eqnarray}
where $R(r)$ has been defined in~\eref{eq:AphiIDGaussGen} and $C_{10}$ denotes the amplitude.
We solve~\eref{eq:DivEIDGauss} with boundary conditions $\lim_{r\rightarrow\infty} \hE^{i} =0$ to find 
\begin{eqnarray}
\label{eq:SolGaussIDl1m0}
\fl
\hE^{r} = \frac{ C_{10} \mu^{2}_{\rm V} \sigma}{\sqrt{48\pi}} \frac{\cos\theta}{r^3}
        \Big\{
                2 \sigma \left( r^2 + r r_{0} + r^{2}_{0} + \sigma^{2} \right) R(r)
              - 2 \sigma \left( r^{2}_{0} + \sigma^{2} \right) \exp\left[ -\frac{r^{2}_{0}}{\sigma^{2}} \right]
\nonumber \\
              - \sqrt{\pi} \left( r^{3} + 2 r^{3}_{0} + 3 r_{0} \sigma^{2} \right) \erf\left[\frac{ r-r_{0} }{\sigma}\right]
\nonumber \\
              - \sqrt{\pi} r_{0} \left( 2 r^{2}_{0} + 3 \sigma^{2} \right) \erf\left[\frac{r_{0}}{\sigma} \right]
              + \sqrt{\pi} r^{3}
        \Big\}
\,,\\
\fl
\hE^{\theta} = \frac{ C_{10} \mu^{2}_{\rm V} \sigma}{\sqrt{192\pi}} \frac{\sin\theta}{r^4}
        \Big\{
                2 \sigma \left( r^2 + r r_{0} + r^{2}_{0} + \sigma^{2} \right) R(r)
              - 2 \sigma \left( r^{2}_{0} + \sigma^{2} \right) \exp\left[ -\frac{r^{2}_{0}}{\sigma^{2}} \right]
\nonumber \\
              + \sqrt{\pi} \left( 2 r^3 - 2 r^{3}_{0} - 3 r_{0} \sigma^2 \right) \erf\left[\frac{r-r_{0}}{\sigma}\right]
\nonumber \\
              - \sqrt{\pi} r_{0} \left( 2 r^{2}_{0} + 3 \sigma^2 \right) \erf\left[\frac{r_{0}}{\sigma} \right]
              - 2 \sqrt{\pi} r^{3} 
        \Big\} 
\,,\\
\fl
\hE^{\varphi} =  0\,.
\end{eqnarray}
The source terms in the Hamiltonian constraint~\eref{eq:HamiltonianIDGauss},
given by our solution~\eref{eq:GaussIDl1m0Aphi} and~\eref{eq:SolGaussIDl1m0},
remain regular (and indeed vanish) at the puncture. We show in figure~\ref{fig:u_dipole} an example of the correction term $u$
(see equation~\eref{eq:u-def}) obtained with this construction.

\begin{figure}[tbhp]
\centering
\includegraphics[width=0.45\textwidth]{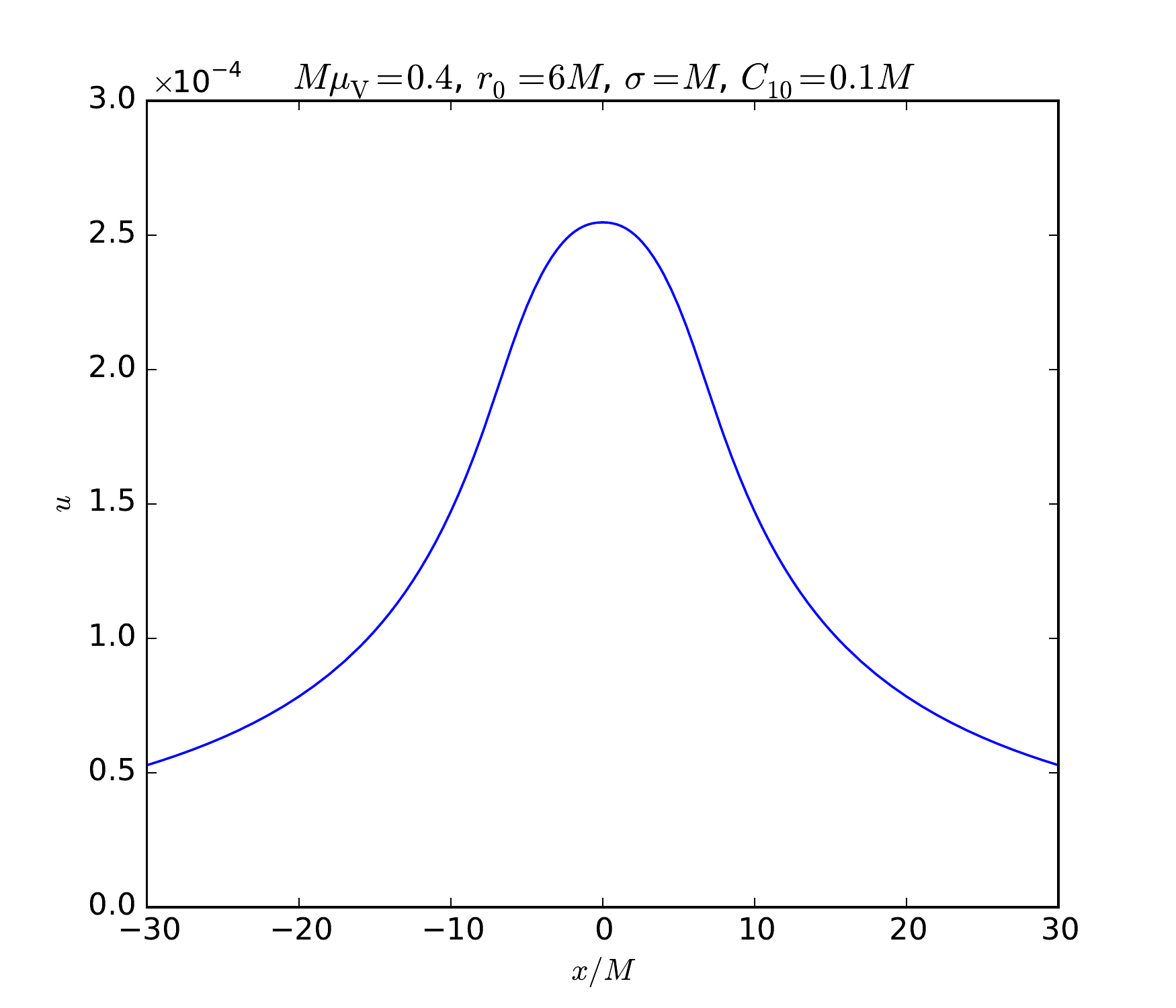}
\includegraphics[width=0.45\textwidth]{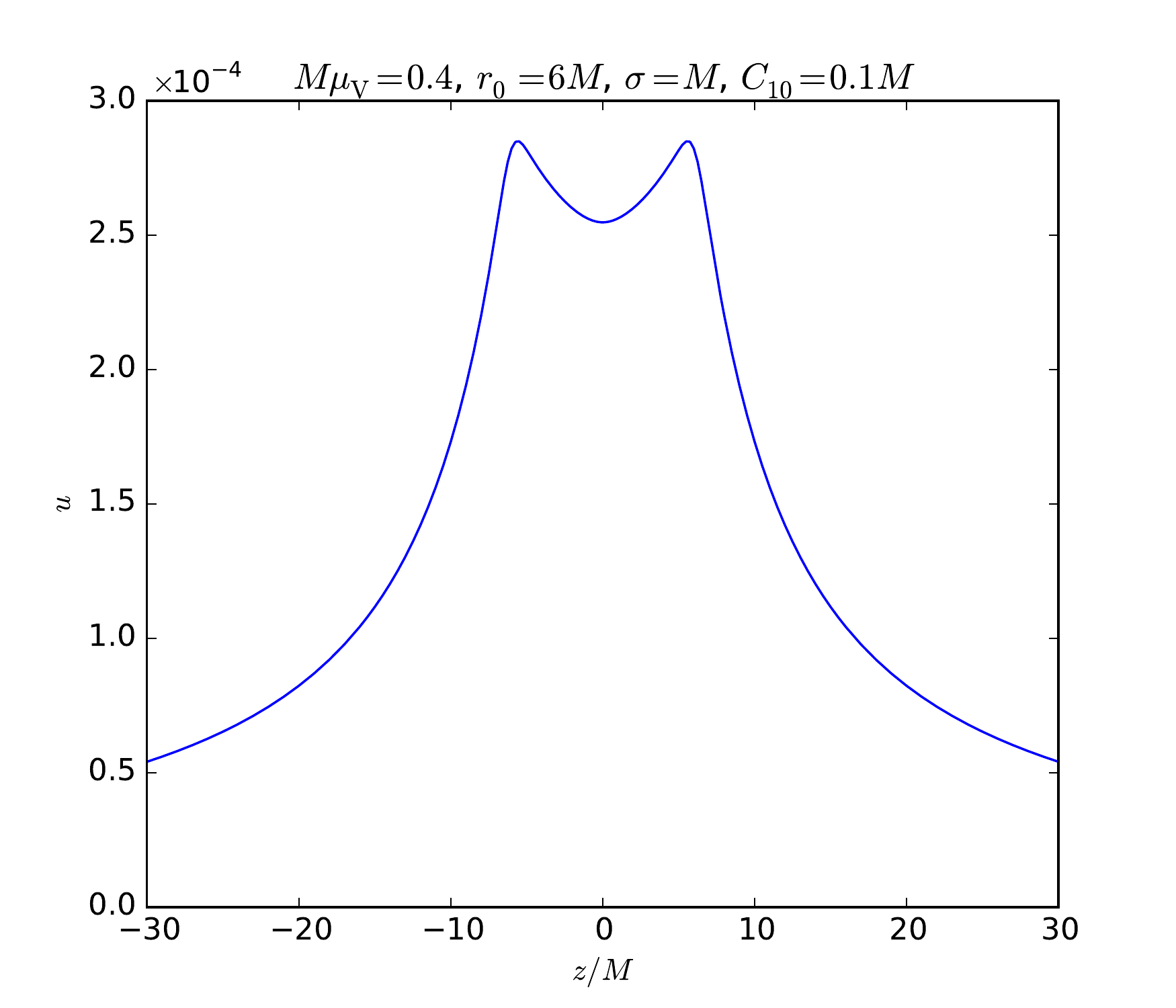}
\caption[]{Typical example of the correction term $u$ for a dipole
  configuration, equation~\eref{eq:GaussIDl1m0Aphi}, plotted along the $x$ (left
  panel) and $z$ (right panel) axis. 
  It refers to a cloud with mass coupling $M\mu_{\rm V}=0.4$ 
  centered around $r_{0}/M=6$ with width $\sigma/M=1$ and amplitude $C_{10}=0.1$.
  All dimensionful quantities are in units of
  the initial BH mass parameter $M=1$. \label{fig:u_dipole} }
\end{figure}

\end{description}

\section{Wave extraction}
\label{sec:WaveExtraction}
To obtain information about waves 
emitted in form of gravitational 
and electromagnetic radiation we employ the Newman-Penrose (NP)
formalism~\cite{Newman:1961qr}.
We will use the notation conventions of~\cite{Alcubierre:2008}.
In this spinor-inspired formalism the radiative degrees of freedom are 
given by a set of (complex) scalars.
They are defined as contractions of the Weyl and Maxwell tensor, respectively, 
with the vectors of a null tetrad $(k^{\mu},l^{\mu},m^{\mu}, \bar{m}^{\mu})$
with $k^{\mu} l_{\mu} = - 1 = -m^{\mu} \bar{m}_{\mu}$.
%
We construct the null-tetrad vectors from
\begin{eqnarray*}
k^{\mu} & =  \frac{1}{\sqrt{2}} \left( n^{\mu} - u^{\mu} \right)
\,,\qquad
l^{\mu} &  =   \frac{1}{\sqrt{2}} \left( n^{\mu} + u^{\mu} \right)
\,,\\
m^{\mu} & = \frac{1}{\sqrt{2}} \left( v^{\mu} + \imath w^{\mu} \right)
\,,\quad
\bar{m}^{\mu} & = \frac{1}{\sqrt{2}} \left( v^{\mu} - \imath w^{\mu} \right)
\,,
\end{eqnarray*}
where $n^{\mu}$ is the timelike unit normal vector
and the spatial vectors $(u^{i},v^{i},w^{i})$ 
form a Cartesian orthonormal basis 
(see, e.g.,~(A2)-(A5) in~\cite{Witek:2010qc}).
Asymptotically the basis vectors $(u^{i},v^{i},w^{i})$ behave as the unit radial, polar and azimuthal vectors.

We extract the radiation carried in the vector fields
by computing the NP scalars
%
\begin{eqnarray}
\label{eq:PhiEMDef}
\Phi_{0} & = & W_{\mu\nu} l^{\mu} m^{\nu}
\,,\quad
\Phi_{1}   =   \frac{1}{2} W_{\mu\nu} \left( l^{\mu} k^{\nu} + \bar{m}^{\mu} m^{\nu} \right)
\,,\quad
\Phi_{2}   =   W_{\mu\nu} \bar{m}^{\mu} k^{\nu}
\,,
\end{eqnarray}
which are reconstructed from the dynamical variables $(\A_{i},E^{i})$ using
\begin{eqnarray}
\Phi_{0} & = &- \frac{1}{2} \left( E_{i} v^{i} + u^{i} v^{j} \left( D_{j}\A_{i} - D_{i} \A_{j} \right) \right) \nonumber \\
          & & - \frac{\imath}{2} \left( E_{i} w^{i} + u^{i} v^{j} \left( D_{j} \A_{i} - D_{i} \A_{j} \right) \right)
\,,\label{eq:Phi0}\\
\Phi_{1} & = &  \frac{1}{2} E_{i} u^{i} 
             - \frac{\imath}{2} v^{i} w^{j} \left( D_{j} \A_{i} - D_{i} \A_{j} \right)
\,,\label{eq:Phi1}\\
\Phi_{2} & = & \frac{1}{2} \left( E_{i} v^{i} - u^{i} v^{j} \left( D_{j}\A_{i} - D_{i}\A_{j} \right) \right) \nonumber \\
           & & - \frac{\imath}{2} \left( E_{i} w^{i} - u^{i} w^{j} \left( D_{j}\A_{i} - D_{i}\A_{j} \right) \right)
\,.\label{eq:Phi2}
\end{eqnarray}
%

The gravitational radiation is encoded in the in- and outgoing Weyl scalars
(in an appropriate null tetrad)
\begin{eqnarray}
\label{eq:Psi0Psi4Def}
\Psi_{0}  =  C_{\mu\nu\rho\sigma} k^{\mu} \bar{m}^{\nu} k^{\rho} \bar{m}^{\sigma}
\,,\qquad
\Psi_{4} =   C_{\mu\nu\rho\sigma} l^{\mu} m^{\nu} l^{\rho} m^{\sigma}
\,.
\end{eqnarray}
In analogy with electromagnetism it is convenient to decompose the Weyl tensor
into its electric and magnetic components
\begin{eqnarray}
\label{eq:WeylEB}
E_{ij} = & \gamma^{\mu}{}_{i} \gamma^{\rho}{}_{j} C_{\mu\nu\rho\sigma} n^{\nu} n^{\sigma}
\,,\qquad
B_{ij} =   \gamma^{\mu}{}_{i} \gamma^{\rho}{}_{j} \,^{\ast} C_{\mu\nu\rho\sigma} n^{\nu} n^{\sigma} 
\,,
\end{eqnarray}
where $^{\ast} C_{\mu\nu\rho\sigma} = \frac{1}{2} \epsilon^{\alpha\beta}{}_{\rho\sigma} C_{\mu\nu\alpha\beta}$
denotes the dual Weyl tensor. 
In terms of the evolution variables they are given by
\begin{eqnarray}
\label{eq:WeylBinAij}
B_{ij} & = & \epsilon_{(i|}{}^{kl} D_{k} A_{|j)l}
\,,\\
\label{eq:WeylEinAij}
E_{ij} 
       & = & R^{\rm TF}_{ij} - [A_{ik} A^{k}{}_{j}]^{\rm TF} + \frac{1}{3} K A_{ij} 
\nonumber \\ & &
        + \left[ E_{i} E_{j} - \mu^{2}_{\rm V} \A_{i} \A_{j}
                - \gamma^{kl} \left(D_{k}\A_{i} - D_{i}\A_{k} \right)
                              \left(D_{l}\A_{j} - D_{j}\A_{l} \right)
        \right]^{\rm TF}\,,
\end{eqnarray}
where $A_{ij}$ is the trace-free part of the extrinsic curvature,
$[.]^{\rm TF}$ denotes the tracefree part (computed with the physical metric $\gamma_{ij}$)
and in the last relation we used~\eref{eq:dtKij}.
Finally, the Weyl scalars are computed from
\begin{eqnarray}
\label{eq:Psi0Psi4inEB}
\Psi_{0} & = & 
          \frac{1}{2} E_{ij} \left( v^{i} v^{j} - w^{i} w^{j} \right) + B_{ij} v^{i} w^{j} 
\nonumber \\
        & & + \imath \left( 
                  E_{ij} v^{i} w^{j} 
                - \frac{1}{2} B_{ij} \left( v^{i} v^{j} - w^{i} w^{j} \right)
          \right)
\,,\\
\Psi_{4} & = &
          \frac{1}{2} E_{ij} \left( v^{i} v^{j} - w^{i} w^{j} \right) - B_{ij} v^{i} w^{j}
\nonumber \\
        & & - \imath \left(
                  E_{ij} v^{i} w^{j}
                + \frac{1}{2} B_{ij} \left( v^{i} v^{j} - w^{i} w^{j} \right)
          \right)
\,.
\end{eqnarray}
in terms of the electric and magnetic parts of the Weyl tensor in the Cartesian orthonormal 
basis $(u^{i}, v^{i}, w^{i})$.

At a given extraction radius $R_\mathrm{ex}$, we perform a multipolar
decomposition by projecting $\Psi_4$, $\Phi_1$ and $\Phi_2$ onto spin-weighted spherical
harmonics with $s=-2$, $0$, and $-1$, respectively,
\begin{eqnarray}
  \Psi_4(t, \theta, \phi) & =
      \sum_{l,m} \psi^{lm}(t)\, _{-2}Y_{lm}(\theta,\phi) \ , \\
  \Phi_1(t, \theta, \phi) & =
      \sum_{l,m} \phi_{1}^{lm}(t)\, _{0}Y_{lm}(\theta,\phi) \ ,
      \label{eq:multipole_Phi1} \\
  \Phi_2(t, \theta, \phi) & =
      \sum_{l,m} \phi_{2}^{lm}(t)\, _{-1}Y_{lm}(\theta,\phi) \ .\label{eq:multipole_Phi2}
\end{eqnarray}

\section{Numerical results}
\label{sec:results}

\subsection{Numerical setup and code description}
\label{ssec:code}
To simulate the interplay between BHs and massive vector fields we have implemented new thorns as part
of the {\textsc{Lean}} code, originally presented in~\cite{Sperhake:2006cy} for vacuum spacetimes. 
\textsc{Lean} uses the BSSN formulation of the Einstein
equations~\cite{Shibata:1995we,Baumgarte:1998te} with the moving puncture
method~\cite{Campanelli:2005dd,Baker:2005vv}, is based on the \textsc{Cactus}
Computational toolkit~\cite{Cactuscode:web}, the \textsc{Carpet} mesh refinement
package~\cite{Schnetter:2003rb,CarpetCode:web} and uses \textsc{AHFinderDirect}
for tracking apparent horizons~\cite{Thornburg:2003sf,Thornburg:1995cp}.
We employ the method-of-lines, where spatial derivatives are approximated by
fourth-order finite difference stencils, and we use the $4^{\rm{th}}$-order Runge-Kutta scheme
for the time integration.
For further details on the
numerical methods we refer the reader to~\cite{Sperhake:2006cy,Cardoso:2014uka}.

The required extensions to \textsc{Lean} to accommodate for the Einstein-Proca system 
were coded \emph{independently} by two of the authors (H.W.\ and~M.Z.), thus allowing for independent cross-checks. 
Furthermore, since one recovers the standard Einstein-Maxwell case 
by setting $\mu_{\rm V} = 0$ in the initial data construction outlined in section~\ref{ssec:ID-analytic},
we have checked $\mu_{\rm V} = 0$ evolutions against the corresponding (and tested) results obtained 
in~\cite{Zilhao:2012gp,Zilhao:2014wqa}. In all cases we have observed excellent agreement.

We have evolved different initial profiles, 
as outlined in section~\ref{sec:init-data},
and monitored gauge invariant quantities such as the area of the BH's apparent horizon~(AH) 
and the NP scalars $\Phi_{0,1,2}$, cf.~(\ref{eq:PhiEMDef}). 
In table~\ref{tab:runsEi} we list the simulations done with the initial configuration outlined in section~\ref{ssec:ID-analytic} where for fixed BH mass parameter $M=1$ we can vary two parameters, $C$ and $\mu_{\rm V}$. Analogously, we list in table~\ref{tab:runsAphi} simulations done with the initial configurations of section~\ref{ssec:ID-Aphi}, where we have the parameters $C_{00}$, $C_{10}$, $\mu_{\rm V}$, $r_0$ and $\sigma$.

To conclude the description of the implementations, we summarize their accuracy. 
We have simulated SID with a mass coupling $M\mu_{\rm V}=0.4$ and amplitude $C/M=0.4$ at three different resolutions
$h_{\rm c} / M =1/84$, $h_{\rm m} / M = 1/92$ and $h_{\rm h} / M = 1/100$.
A convergence analysis reveals a numerical error in the waveforms $\phi^{00}_{1}$ of 
$\Delta\phi^{00}_{1}/\phi^{00}_{1,h} \lesssim 3\%$ at early times ($t/M=200$)
which increases to $\Delta\phi^{00}_{1}/\phi^{00}_{1,h} \lesssim 9\%$ after an evolution of $t/M=2000$.

\begin{table}[tbhp]
  \centering
  \caption{\label{tab:runsEi}
    List of simulations performed with the SID of
    section~\ref{ssec:ID-analytic}, along with parameters used. $M$ always denotes the mass 
    parameter in the initial data construction. The numerical grid
    structure used (in the notation of section~II~E of~\cite{Sperhake:2006cy}) was the
    following $\{(256,176,64,32,16,8,2,1,0.5), M/80\}$.
    Note, that we have actually run a larger set of simulations and present here only those that are shown later on.
  }
  
\begin{tabular*}{\textwidth}{@{\extracolsep{\fill}}lcc}
\br
Run                   &           $C/M$ &    $M \mu_{\rm V}$   \\
\mr
%
\verb|Ei_c=0.4_mu=0.2|  &          0.4    &        0.2   \\ 
\verb|Ei_c=0.05_mu=0.4| &         0.05    &        0.4   \\ 
\verb|Ei_c=0.4_mu=0.4|  &          0.4    &        0.4   \\ 
%
\verb|Ei_c=0.4_mu=0.5|  &          0.4    &        0.5   \\ 
%
\verb|Ei_c=0.4_mu=0.9|  &          0.4    &        0.9   \\ 
%
%
\verb|Ei_c=0.4_mu=1.5|  &          0.4    &        1.5   \\ 
\br
\end{tabular*}

\end{table}

\begin{table}[tbhp]
  \centering
  \caption{\label{tab:runsAphi}
    List of simulations performed with the GID of
    section~\ref{ssec:ID-Aphi}, along with parameters used. 
    As before $M=1$ is the initial mass parameter of the BH.
    The numerical grid structure used (in the notation of section~II~E of~\cite{Sperhake:2006cy}) 
    is given by $\{(256,176,64,32,16,8,2,1,0.5), M/80\}$.
  }
  
\begin{tabular*}{\textwidth}{@{\extracolsep{\fill}}lcccccc}
\br
Run                        & $C_{00}/M$ & $C_{10}/M$ & $r_{0}/M$ & $\sigma/M$ & $\mu_{\rm V} M$ 
& $M_{\rm{ADM}}/ M$ \\
\mr
\verb|phi_c00=0.1_mu=0.2|       &  0.1       &  0         &  6.0     & 1.0         & 0.2     & 1.00027 \\ 
\verb|phi_c00=1.0_mu=0.2|       &  1.0       &  0         &  6.0     & 1.0         & 0.2     & 1.02668 \\
\verb|phi_c00=0.1_mu=0.4_r0=5|  &  0.1       &  0         &  5.0     & 1.0         & 0.4     & 1.00213 \\
\verb|phi_c00=0.1_mu=0.4_r0=12| &  0.1       &  0         & 12.0     & 1.0         & 0.4     & 1.00529 \\
\hline
\verb|phi_c10=0.1_mu=0.1| &  0         &  0.1       &  6.0     & 1.0         & 0.1     & 1.00011 \\
\verb|phi_c10=1.0_mu=0.1| &  0         &  1.0       &  6.0     & 1.0         & 0.1     & 1.01091 \\ 
\verb|phi_c10=0.1_mu=0.2| &  0         &  0.1       &  6.0     & 1.0         & 0.2     & 1.00051 \\
%
%
\verb|phi_c10=0.1_mu=0.4| &  0         &  0.1       &  6.0     & 1.0         & 0.4     &   1.00320\\ 
\br
\end{tabular*}

\end{table}

\subsection{Evolutions of Proca fields coupled to nonrotating BHs}

\begin{figure}[htbp!]
\centering
\includegraphics[width=0.60\textwidth]{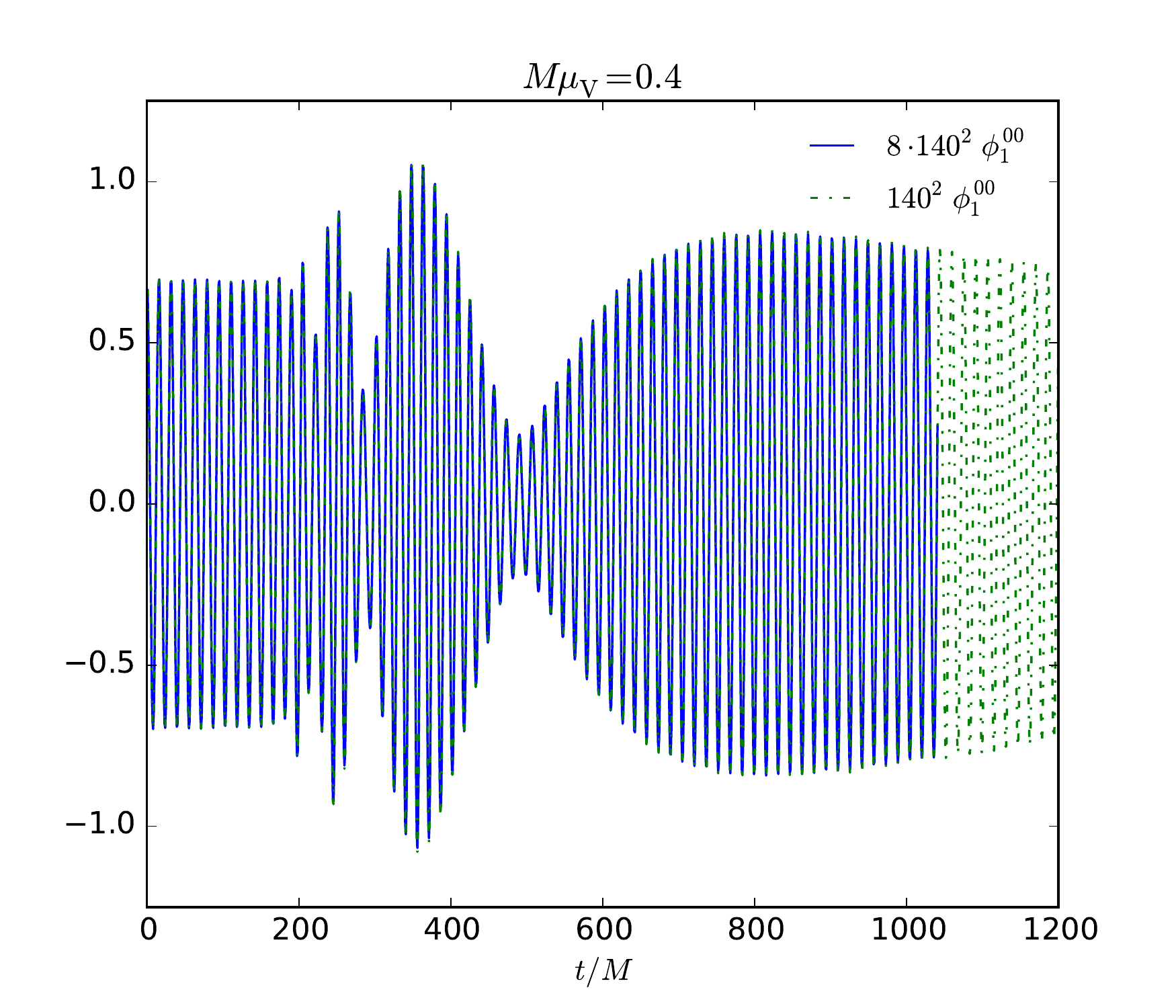}
\caption[]{$\phi_1^{00}$ for the SID configurations of
  section~\ref{ssec:ID-analytic} with $M\mu_{\rm V}=0.4$, extracted at $R_{\rm ex}=140M$. 
  The (blue) solid line shows the case with amplitude $C=0.05$,  
  while the (green) dashed-dotted line refers to an amplitude $C=0.4$.
  We have scaled the $C=0.05$ results by $8=0.4/0.05$; the two curves then overlap, 
  confirming that our runs still lie in the linear regime.
  The beating patterns in these curves are due to the presence of multiple overtones excited by the initial data.
   \label{fig:phi1_compare}}
\end{figure}

\begin{figure}[htpb!]
\centering
\includegraphics[width=0.45\textwidth]{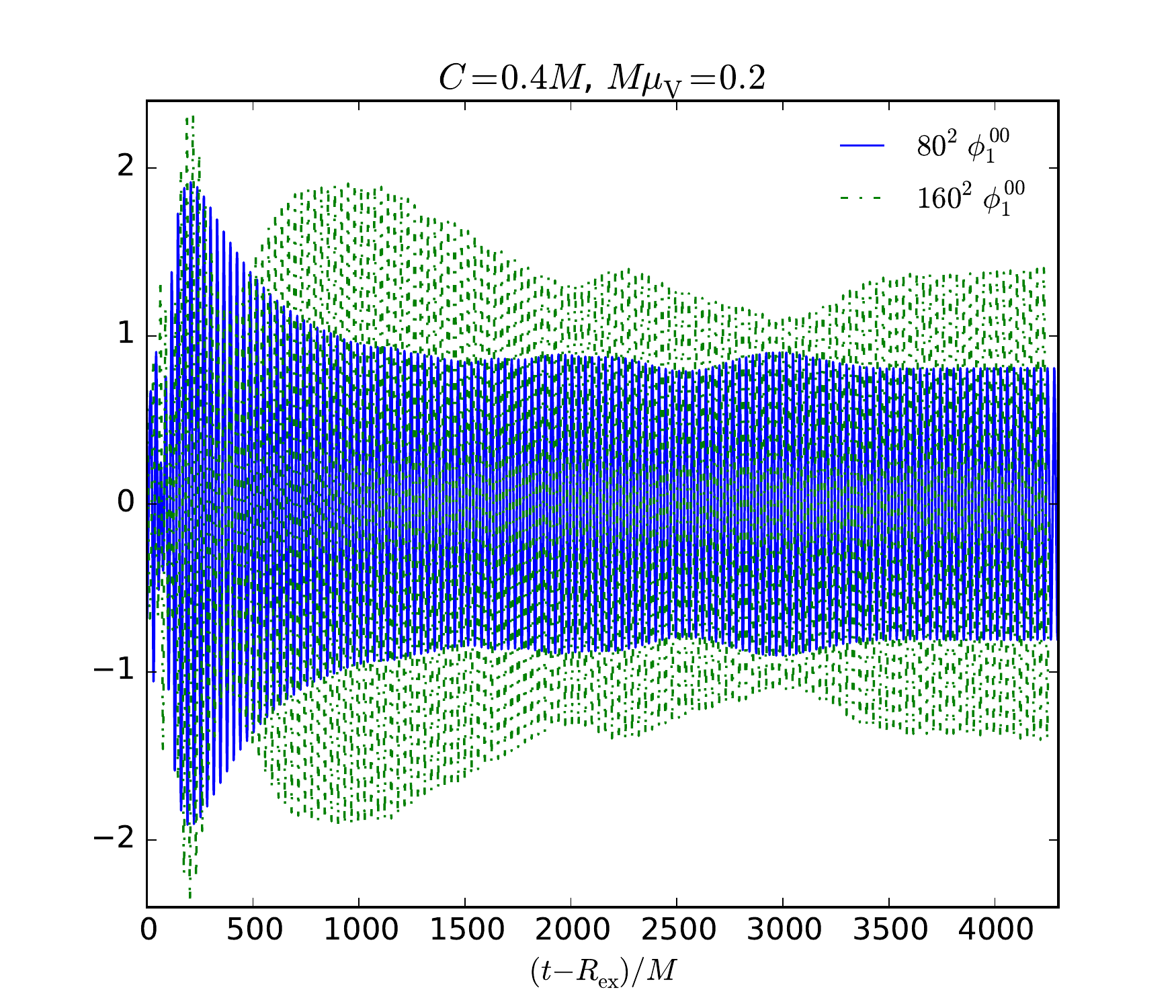}
\includegraphics[width=0.45\textwidth]{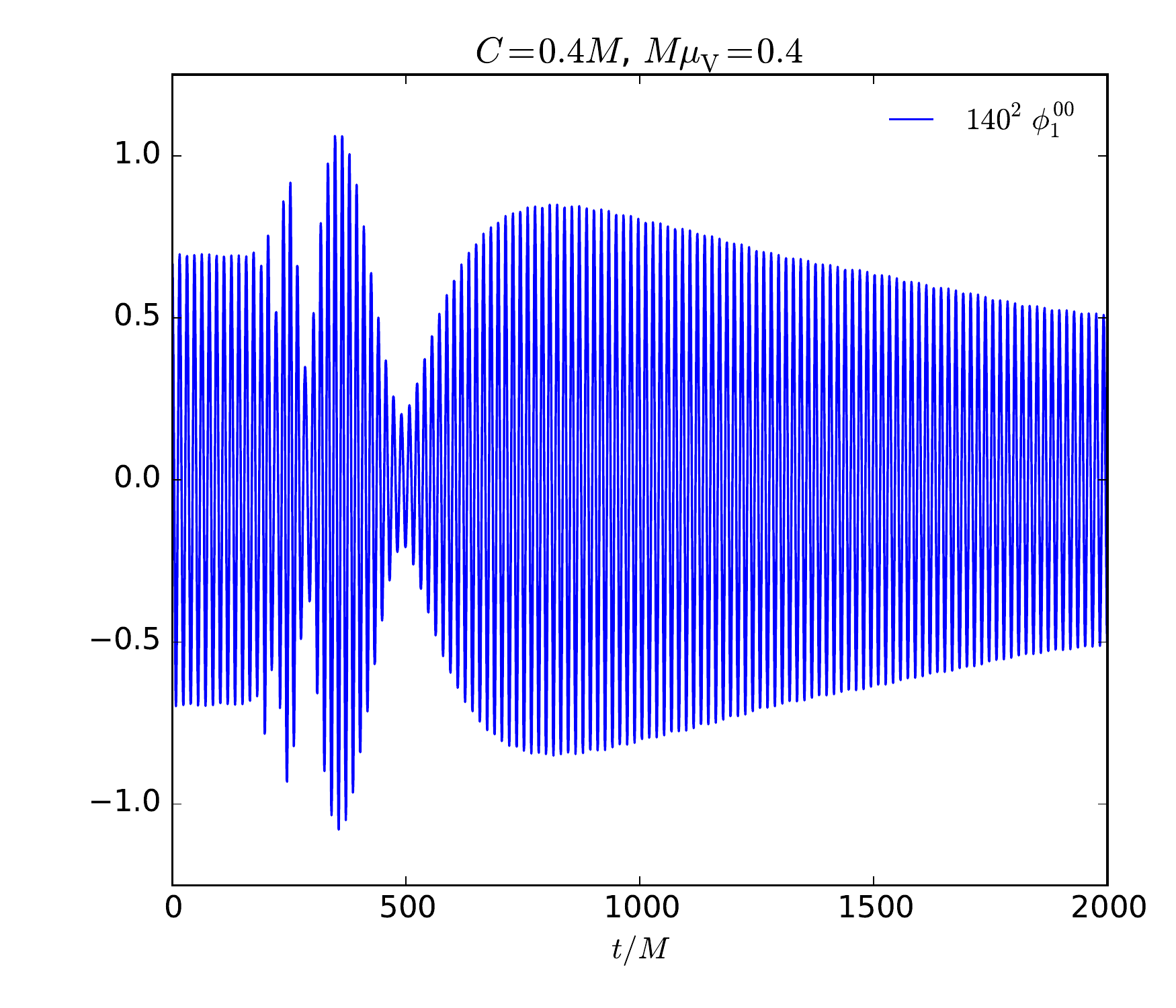} \\
\includegraphics[width=0.45\textwidth]{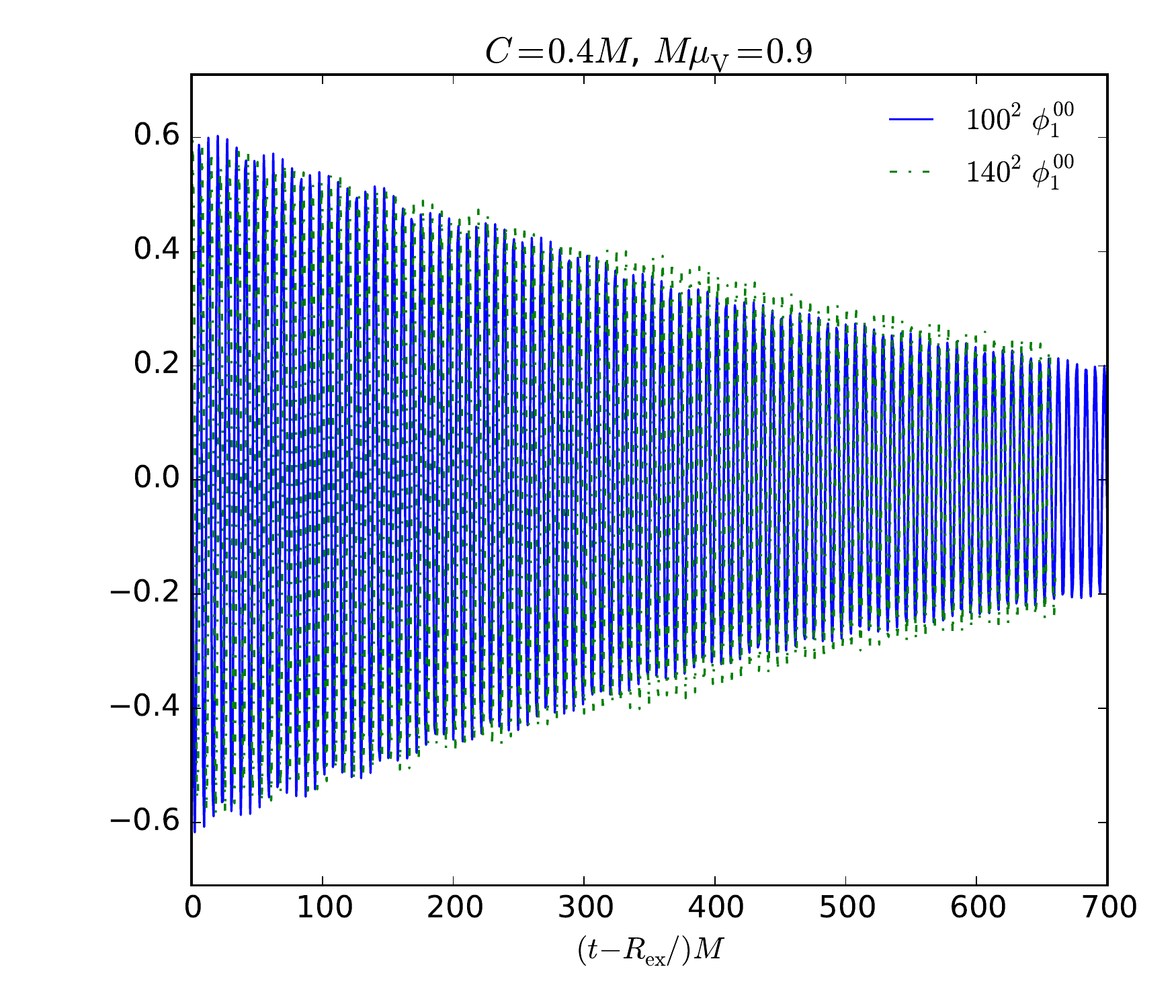} 
\includegraphics[width=0.45\textwidth]{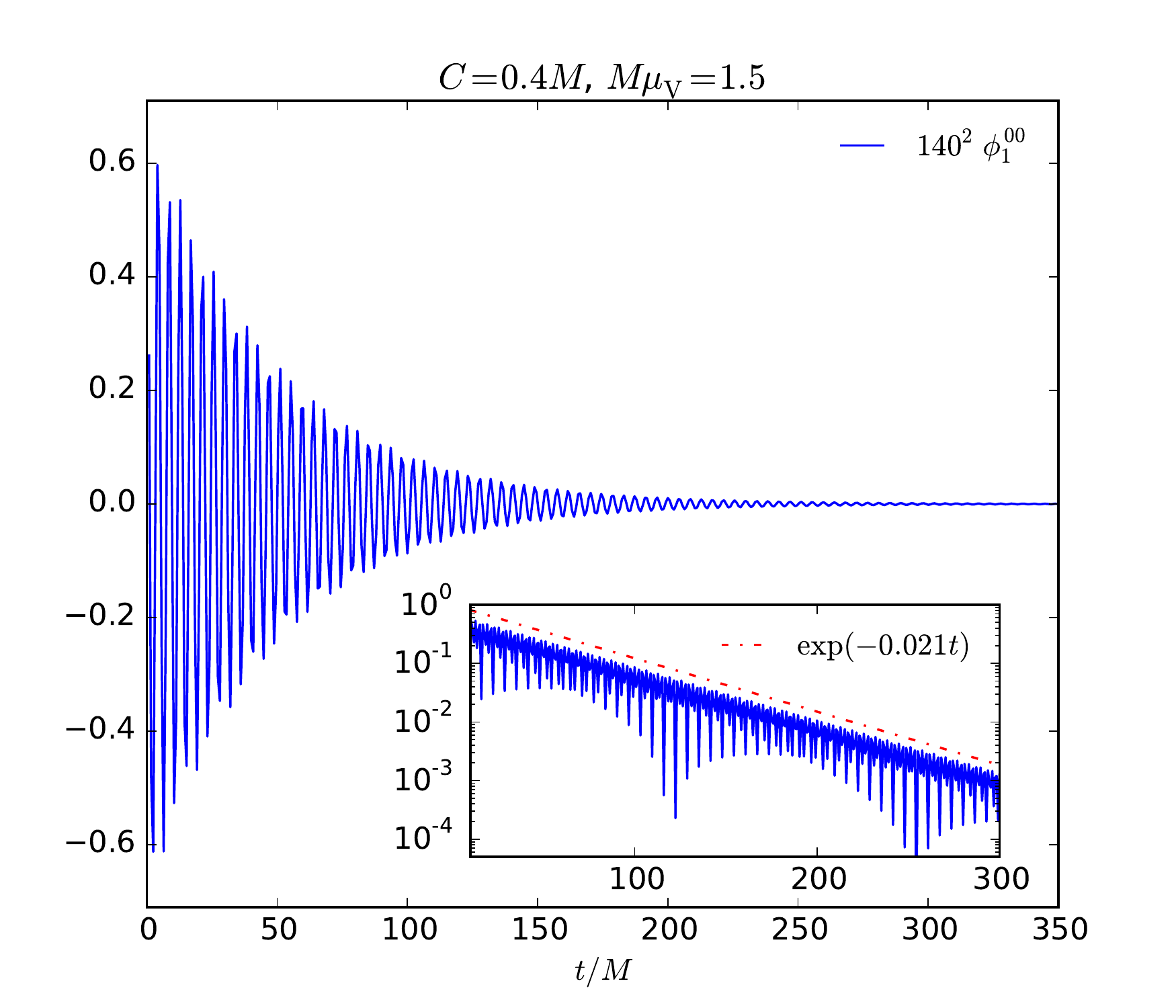}
\caption[]{Results for SID configurations as described in section~\ref{ssec:ID-analytic} with  amplitude $C=0.4M$
        and different mass-couplings, varying between $M\mu_{\rm V}=0.2$ (top left) to $M\mu_{\rm V}=1.5$ (bottom right).
        We plot the monopole waveform $\phi_{1}^{00}$, rescaled by the respective extraction radius.
        Due to spherical symmetry this is the only nontrivial mode excited throughout the evolution.
        In a massless $\mu_{\rm V}=0$ configuration, $\phi_1^{00}$ would measure the
        monopole component of the electric field, i.e., the electric charge. 
        The inset in the last panel shows the data in a logarithmic scale with a fit of the form $e^{-0.021t}$, which matches very well with  frequency-domain calculations.
        \label{fig:phi1}}
\end{figure}

\begin{figure}[htbp]
\centering
\includegraphics[width=0.6\textwidth]{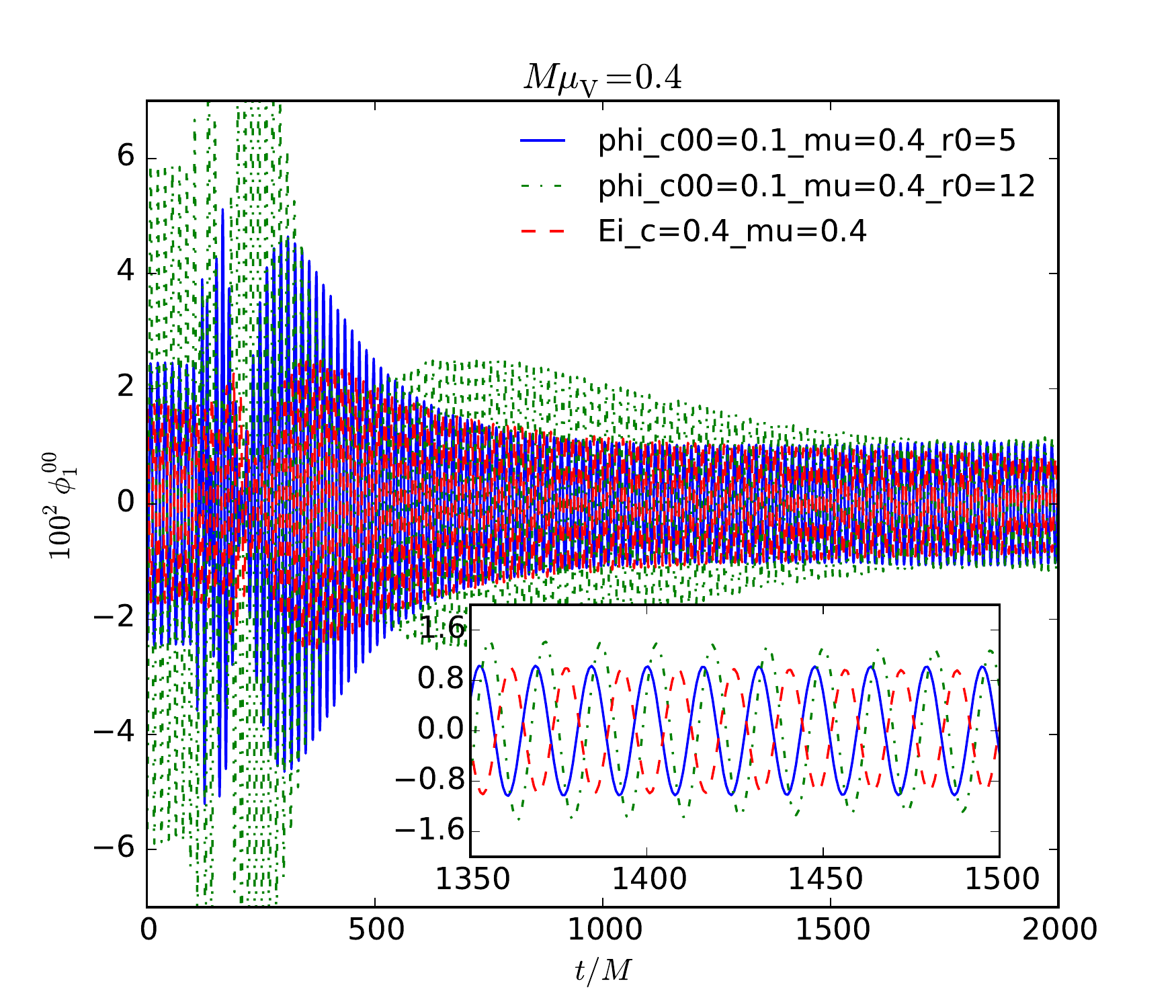}
\caption[]{We present the monopole mode $\phi_{1}^{00}$, rescaled by the extraction radius $R_{\rm ex}=100 M$,
           for a Proca field with $M\mu_{\rm V}=0.4$. The different curves correspond to different initial configurations;
           specifically the (blue) solid and (green) short-dashed lines are type I GID (see section~\ref{ssec:ID-Aphi}) where
           the spherically symmetric Gaussian was centered around $r_{0}=5M$ and $r_{0}=12M$, respectively,
           and the (red) long-dashed line refers to SID (see section~\ref{ssec:ID-analytic}).
           Although the BH's immediate response varies substantially for different initial setups 
           yielding more or less pronounced beating patterns, the overall oscillation frequency appears to be the same
           as can be seen in the inset.
           \label{fig:phi100_Aphi}}
\end{figure}

\begin{figure}[htpb!]
\centering
\includegraphics[width=0.45\textwidth]{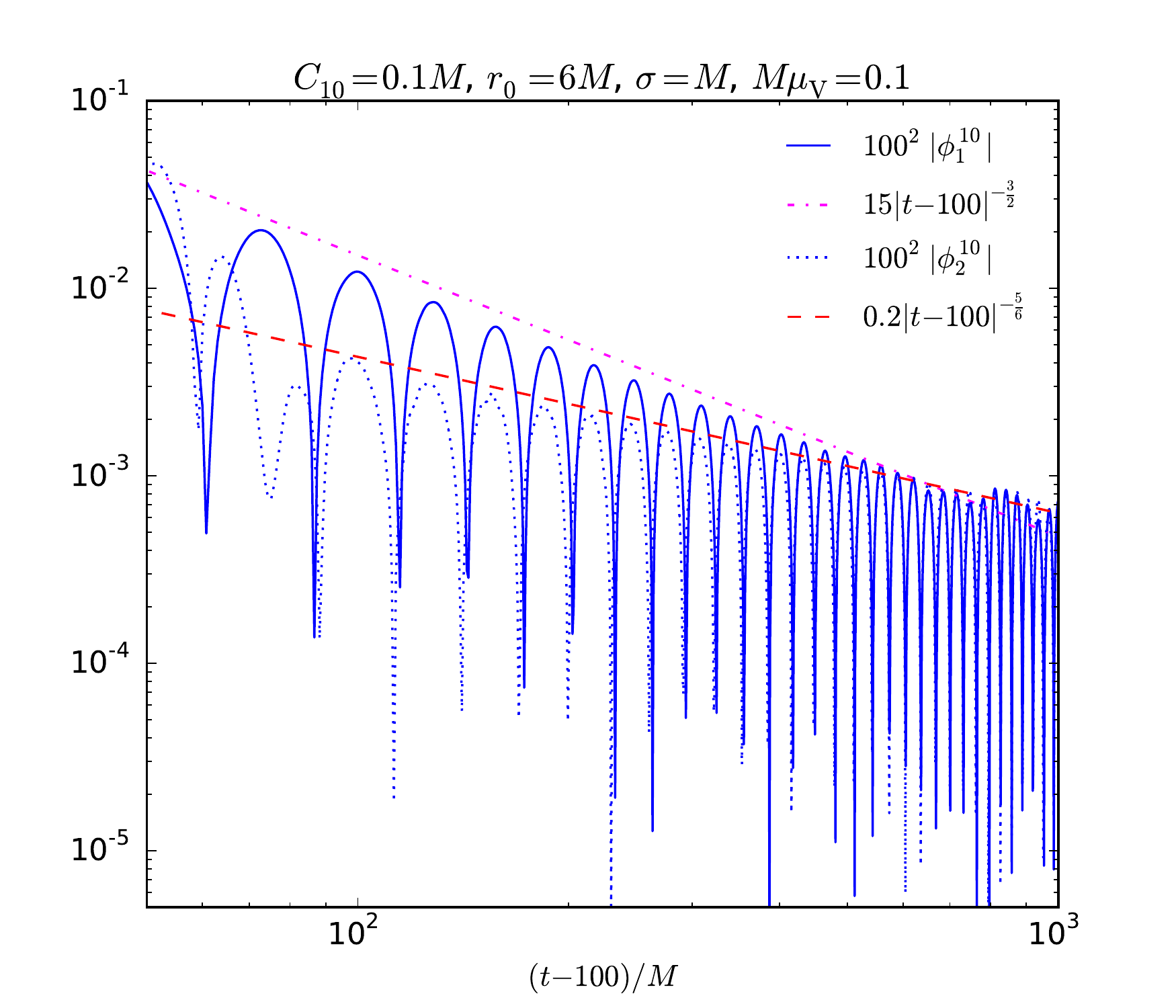}
\includegraphics[width=0.45\textwidth]{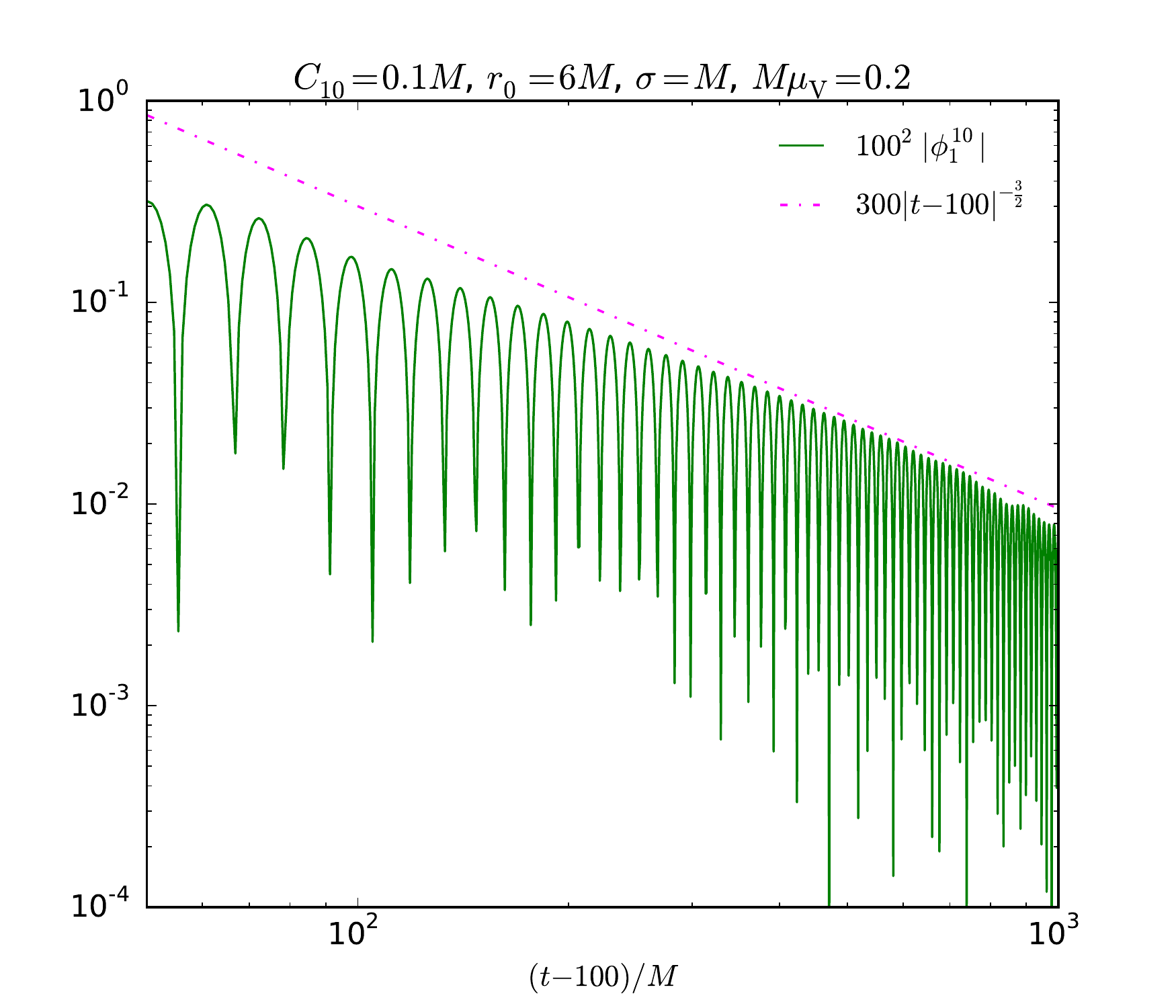} \\
\includegraphics[width=0.45\textwidth]{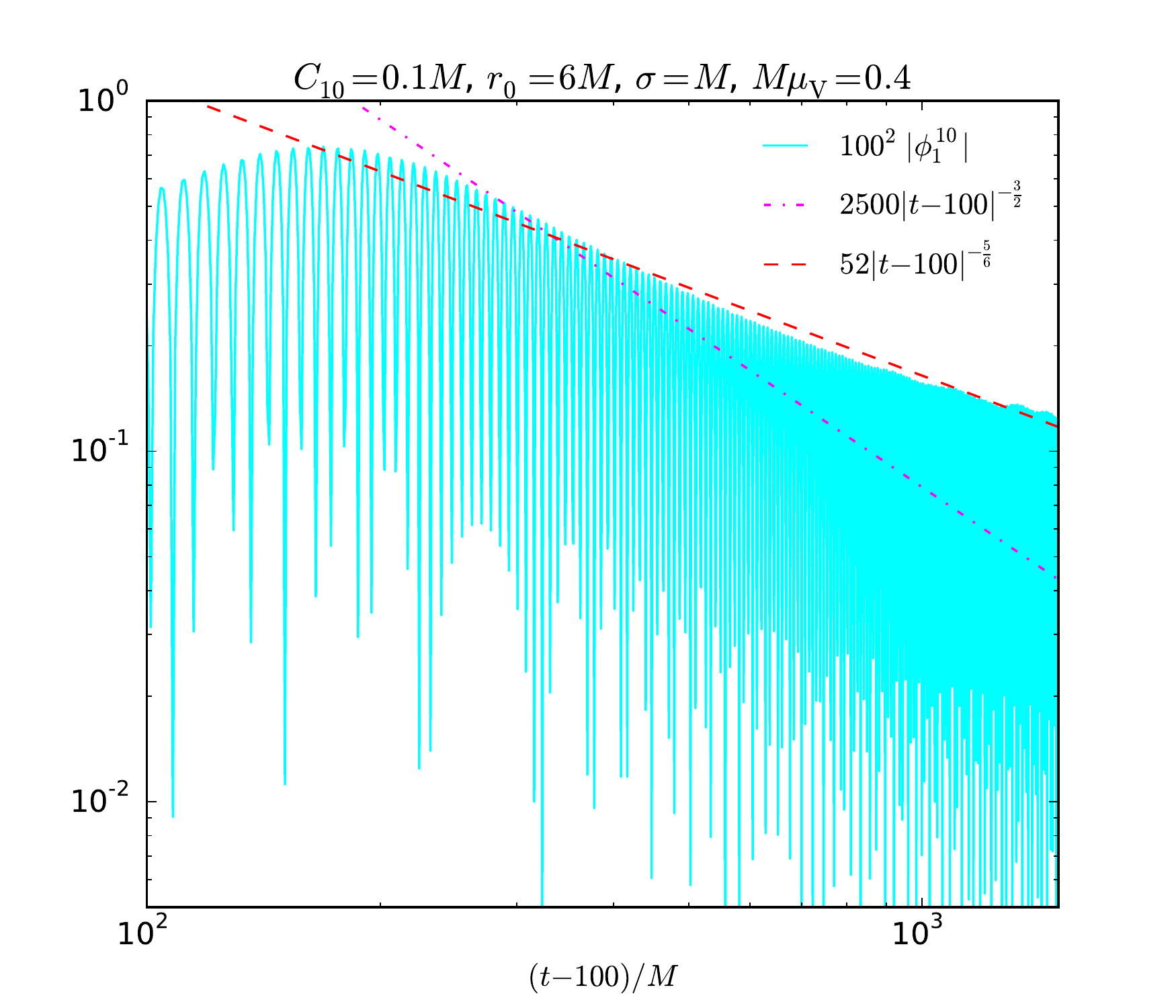}
\includegraphics[width=0.45\textwidth]{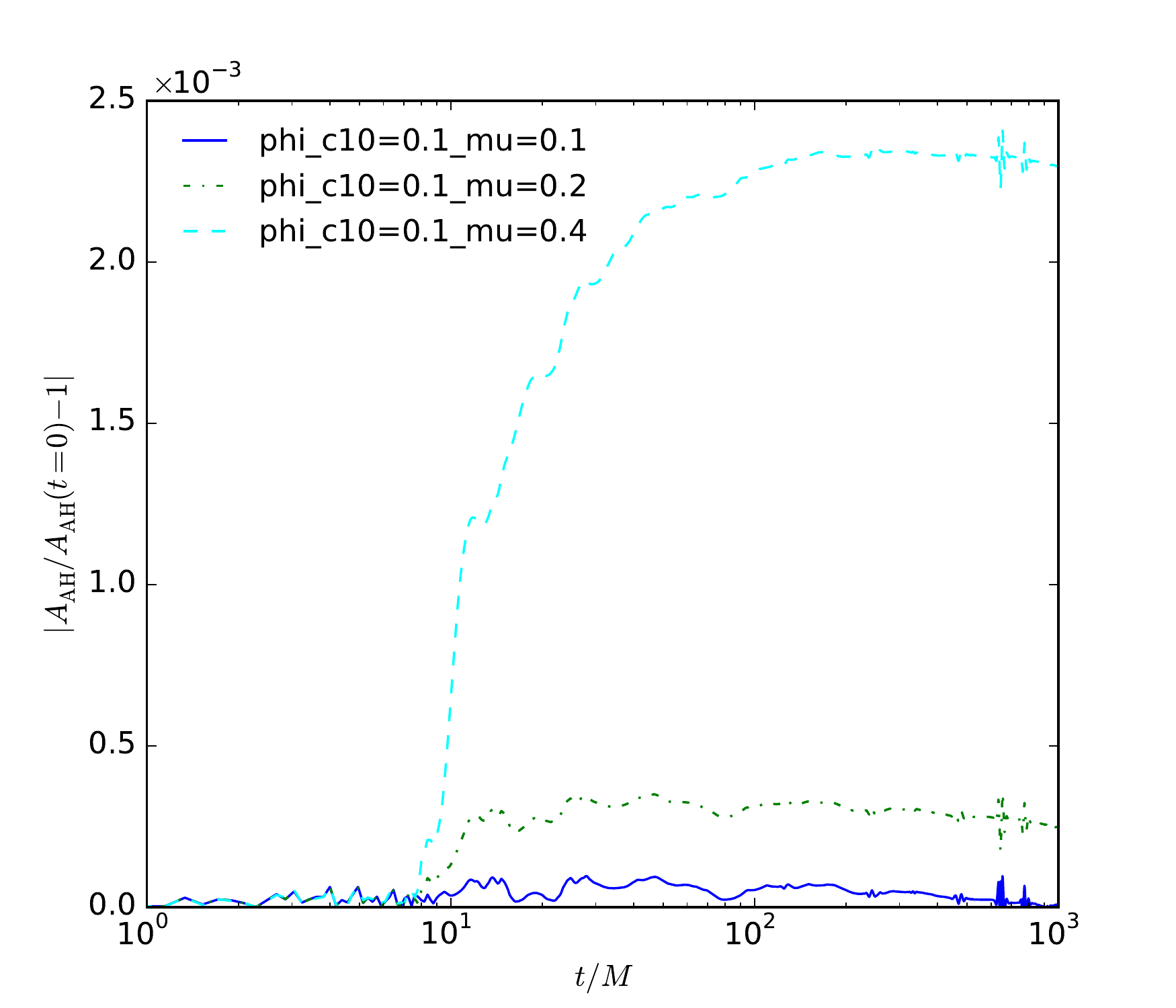}
\caption[]{We present $\phi^{10}_{1}$ (solid lines) and $\phi^{10}_{2}$ (dotted
  line in the top left panel) extracted at $R_{\rm ex}=100M$, obtained from an
  initial GID dipole configuration
  with $M\mu_{\rm V}=0.1$ (top left), $M\mu_{\rm V}=0.2$ (top right) and
  $M\mu_{\rm V}=0.4$ (bottom left).  The (magenta) short-dashed and (red)
  long-dashed lines are envelopes of the (massive) power-law tails corresponding,
  respectively, to intermediate-time tails $\sim t^{-3/2}$ compatible with an
  $S=-1,l=1$ excitation, and to very late-time tails $\sim t^{-5/6}$.  In the
  bottom right panel we plot the relative change in the area of the corresponding
  AHs. The AH area increases during the first interaction between field and BH,
  and saturates to an approximately constant value when the power-law decay stage
  is reached. The downward trend observed at late times in the AH area is likely
  due to constraint violating modes, possibly due to reflections at the outer boundary of
  the computational domain (sitting at $\sim 256M$).
  \label{fig:AHarea}}
\end{figure}

Our numerical results are summarized in figures~\ref{fig:phi1_compare}--\ref{fig:AHarea}
and the generic, qualitative features agree with previous studies on linearized Proca fields around BHs~\cite{Witek:2012tr} and 
with nonlinear evolutions of massive scalars~\cite{Okawa:2014nda}.

For example, figure~\ref{fig:phi1_compare} shows the evolution of SID data, with two different---and small---values of the amplitude $C$, for $M\mu_{\rm V}=0.4$.
The two curves overlap after a linear re-scaling, showing that nonlinear effects are negligible for this data. More importantly,
the time evolution generically shows the presence of a beating pattern. As explained previously~\cite{Witek:2012tr,Okawa:2014nda},
these are the result of the presence of multiple overtones with similar frequencies 
(i.e., modes with different $n$ in equation~\eref{hydrogen_spectra_vector}).

Similar results are obtained for other spherically symmetric configurations using SID data. 
In figure~\ref{fig:phi1} we show the scalar profile $\phi_1^{00}$ 
for various mass couplings and keeping the amplitude $C$ of the initial data fixed.
In all cases we observe a (slowly) damped oscillatory pattern, corresponding to QBS-excitations 
(see section~\ref{ssec:LiteratureSummary}). Oscillation frequencies and damping times are here uniquely 
determined by the mass parameter $\mu_{\rm V}$. 
For example, for $M\mu_{\rm V}=1.5$ the waveform is best fitted by a damped sinusoid of the form $e^{-0.021t}\sin 1.48t$. 
This agrees very well with the fourth-overtone of Proca fields around a nonrotating BH, which we computed using frequency-domain calculations~\cite{Berti:2009kk,Rosa:2011my,Pani:2012bp} to
be $\omega=1.476-\imath 0.02158$.
For smaller mass couplings, the signal is extremely long-lived, as can be seen from the top panel in
figure~\ref{fig:phi1}.

%

In fact, very long-lived, {\it spherically symmetric} configurations are one of the features borne out of our work.
They arise for generic initial data. For example, GID data of section~\ref{ssec:ID-Aphi}, with more free parameters
gives us a richer structure in the oscillation pattern, but also long-lived signals.
This is more clearly seen in figure~\ref{fig:phi100_Aphi}, where we plot $\phi_1^{00}$ 
generated from SID and spherically symmetric GID initial configurations.
The features worth mentioning are the beating pattern due to excitation of different overtones, 
but most specially the late time behaviour, which is independent of the initial data and is, 
basically, a very long-lived damped sinusoid. The signal can last for thousands of dynamical timescales,
and the perturbative results indicate that long-lived modes might be present.

In other words, QBS states are apparent in the evolution of the initial data. In some cases, it is possible to also observe 
late-time, oscillating power-law tails. 
While they have been buried under the long-lived QBS modes in the spherically symmetric cases, we do find the massive tails
for dipolar initial configurations.
We therefore focus now on simulations with type II GID data of section~\ref{ssec:ID-Aphi}.
In figure~\ref{fig:AHarea} we present a set of examples 
referring to the runs  
\verb|phi_c10=0.1_mu=0.1|, \verb|phi_c10=0.1_mu=0.2| and \verb|phi_c10=0.1_mu=0.4|,
where the dipole mode has been excited.
Specifically, we display the $l=1,m=0$ mode of the NP scalars $\phi^{10}_{1}$ and, 
for the smallest mass coupling, $\phi^{10}_{2}$
together with the envelopes for the power-law tails given in~\eref{eq:MassiveTails}.
The former set of waveforms is consistent with an intermediate-time, $S=-1,l=1$ massive tail
$\sim t^{-3/2} \sin( \mu_{\rm V} t)$. 
Furthermore, the waveform for the $M\mu_{\rm V}=0.4$ case 
nicely exhibits the transition from the mode-dependent intermediate-time tail 
to the universal late-time tail $\sim t^{-5/6} \sin(\mu_{\rm V} t)$,
as can be seen in the bottom left panel of figure~\ref{fig:AHarea}.
We expect a similar behaviour for the smaller mass-couplings for which this transition occurs after
$t\gg1000 M$,
longer than our evolution time.
Instead, the second NP scalar $\phi^{10}_{2}$ 
that we show exemplary for the small coupling case (see top left panel of figure~\ref{fig:AHarea})
corresponds to a different polarization for which the intermediate-time tail decays faster and,
indeed, we observe an early onset of the universal late-time tail $\sim t^{-5/6} \sin(\mu_{\rm V} t)$.

To complete the picture illustrating the nonlinear interaction between massive vector fields and 
(nonrotating) BHs we have analysed the properties of the apparent horizon.
We show the relative change in the AH area $\left| A_{\rm AH} / A_{\rm AH}(t=0) - 1 \right|$
in the bottom right panel of figure~\ref{fig:AHarea}.
%
This encodes the direct (nonlinear) response of the BH to the impinging field: 
In all models a fraction of the massive vector field is absorbed by the BH, the exact amount of which depends 
both on the initial energy content in the perturbing field and on its life-time.
This can be seen from figure~\ref{fig:AHarea} by comparing the 
development of the models with small mass couplings (top panel) to the one with $M\mu_{\rm V} =0.4$ (bottom left panel).
In the former cases the AH area increases upon the infall of the field after which it stays approximately constant
consistent with the early onset of the power-law decay in the waveform.
In the latter case the Proca field lingers around yielding a longer absorption phase 
and only when it reaches the tail-stage the AH area saturates to an approximately constant value.

\section{Final remarks}
\label{sec:final}

In this work we have introduced a formalism to numerically integrate the Einstein-Proca system, describing General Relativity minimally coupled to a massive vector field, and we have used this formalism to begin to study how BHs evolve dynamically in this theory. Our results agree with perturbation theory expectations and pave the way for future nonlinear evolutions in more complex scenarios. Of particular interest, currently under development, is the study of configurations with rotating BHs. We find that the time-development of generic classes of initial data lead to extremely long-lived, nearly periodic solutions, generalizing the findings of Ref.~\cite{Barranco:2012qs,Barranco:2013rua} to the vector sector.
Truly periodic, asymptotically flat hairy BHs have recently been ruled out~\cite{Alexakis:2015ara}, but our results show that BHs in Proca theories might develop what for all purposes is a hair on timescales extremely large compared to any meaningful dynamical timescale.

\ack

M.Z.\ is supported by grants 2014-SGR-1474, MEC FPA2010-20807-C02-01,
MEC FPA2010-20807-C02-02, CPAN CSD2007-00042 Consolider-Ingenio 2010,
and ERC Starting Grant HoloLHC-306605.
H.W.\ acknowledges financial support provided under
the {\it ERC-2011-StG 279363--HiDGR} ERC Starting Grant.
V.C.\ acknowledges financial support provided under the European
Union's FP7 ERC Starting Grant ``The dynamics of black holes: testing
the limits of Einstein's theory'' grant agreement no. DyBHo--256667,
and H2020 ERC Consolidator Grant ``Matter and strong-field gravity: New frontiers in Einstein's theory'' grant agreement no. MaGRaTh--646597.
We thank the Yukawa Institute for Theoretical Physics at
Kyoto University for hospitality during the YITP-T-14-1 
workshop on ``Holographic vistas on Gravity and Strings.''
This research was supported in part by Perimeter Institute for Theoretical Physics. 
Research at Perimeter Institute is supported by the Government of Canada through 
Industry Canada and by the Province of Ontario through the Ministry of Economic Development 
$\&$ Innovation.
This work was supported by
the NRHEP 295189 FP7-PEOPLE-2011-IRSES Grant,
and by project PTDC/FIS/116625/2010.
Computations were performed on the ``Baltasar Sete-Sois'' cluster at IST,
and on the COSMOS supercomputer in Cambridge, part of DiRAC facility and
funded by the STFC grants ST/H008586/1, ST/K00333X/1.

\appendix

\section{BSSN equations}
\label{app:BSSNeqs}

For completeness, we display here the complete BSSN equations that we numerically integrate.
\begin{eqnarray}
\fl \left(\p_t - \Lie_{\beta} \right) \A_{i} & = 
- \alpha \chi^{-1} \tilde \gamma_{ij} E^j - \alpha \p_i \Aphi - \Aphi \p_i \alpha \\
\fl \left(\p_t - \Lie_{\beta} \right) Z & = 
\alpha \p_i E^i - \frac{3}{2} \alpha \chi^{-1} E^i \p_i \chi + \alpha \mu_{\rm V}^2 \Aphi - \alpha \kappa Z \\
\fl \left(\p_t - \Lie_{\beta} \right) \Aphi & = 
  \alpha K \Aphi - \alpha \tilde \gamma^{ij} \chi \p_j \A_i
+ \alpha \chi \A_i \tilde \Gamma^i 
+ \frac{\alpha}{2} \A_i \tilde \gamma^{ij} \p_j \chi
- \chi \tilde \gamma^{ij} \A_i \p_j \alpha
- \alpha Z \\
\fl  \left(\p_t - \Lie_{\beta} \right) E^i & = 
\alpha K E^i + \alpha \mu_{\rm V}^2 \chi \tilde \gamma^{ij} \A_j
+ \alpha \chi \tilde \gamma^{ij} \p_j Z
+ \chi^2 \tilde \gamma^{ij} \tilde \gamma^{kl} \p_l \alpha \left( \p_j \A_k - \p_k \A_j \right)
\nonumber \\
 & +  \alpha \chi^2 \tilde \gamma^{ij} \tilde \gamma^{kl}
\left(\tilde D_k \p_j \A_l - \tilde D_k \p_l \A_{j} \right) 
+ \frac{\alpha}{2} \chi \tilde \gamma^{ij} \tilde \gamma^{kl}
\left( \p_j \A_l \p_k \chi - \p_k \A_j \p_l \chi \right)
\\
%
\fl     \left( \partial_t -  \mathcal{L}_\beta \right) \tilde \gamma_{ij} & = 
        - 2 \alpha \tilde A_{ij}\, , \\
\fl     \left( \partial_t -  \mathcal{L}_\beta \right) \chi  & = 
        \frac{2}{3} \alpha \chi K\, , \\
\fl     \left( \partial_t -  \mathcal{L}_\beta \right) K & = 
        [\dots] + 4 \pi \alpha (E + S)\, , \\
\fl     \left( \partial_t -  \mathcal{L}_\beta \right) \tilde A_{ij} & = 
        [\dots] - 8 \pi \alpha \left(
          \chi S_{ij} - \frac{S}{3} \tilde \gamma_{ij}
        \right)\, , \\
\fl     \left( \partial_t -  \mathcal{L}_\beta \right) \tilde \Gamma^i & = 
        [\dots] - 16 \pi \alpha \chi^{-1} j^i\, ,
\end{eqnarray}
where $ [\dots] $ denotes the standard right-hand side of the BSSN equations in
the absence of source terms; the
source terms are determined by
\begin{eqnarray}
  E & \equiv n^\alpha n^\beta T_{\alpha\beta} \, , \\
  j_i & \equiv - \gamma_i{}^\alpha n^\beta T_{\alpha \beta} \, , \\
  S_{ij} & \equiv \gamma^\alpha{}_i \gamma^\beta{}_j T_{\alpha \beta}\, , \\
  S & \equiv \gamma^{ij} S_{ij}\, .
\end{eqnarray}

It is convenient to introduce the ``magnetic'' field $B^i$
\begin{eqnarray}
\label{eq:BofA}
\fl B^i = \epsilon^{ijk} D_j \A_{k} 
 = \chi^{3/2} \epsilon_F^{ijk} \left(
\tilde D_j \A_k 
+ \frac{1}{2} \chi^{-1} ( \A_j \p_k \chi + \A_k \p_j \chi
- \tilde \gamma^{mn} \tilde \gamma_{jk} \A_m \p_n \chi )
\right)
\,,
\end{eqnarray}
as short-hand, where $\epsilon_F^{ijk}$ is a tensor density taking the values of $(0,\pm 1)$ as in flat space.
We then have
\begin{eqnarray}
\fl 4\pi(E+S) & = & \chi^{-1} \tilde \gamma_{ij} \left( E^i E^j + B^i B^j \right) 
+ 2\mu_{\rm V}^2 \Aphi^2 \,, \\
\fl 4 \pi \left(
          \chi S_{ij} - \frac{S}{3} \tilde \gamma_{ij}
        \right) & = &
-\tilde \gamma_{ik} \tilde \gamma_{jl} \chi^{-1}  \left( E^k E^l + B^k B^l \right) 
+ \frac{\tilde \gamma_{ij}}{3} \left(E^2 + B^2 \right)
+ \mu_{\rm V}^2 \chi \A_i \A_j \nonumber \\
&{}-& \mu_{\rm V}^2\frac{\tilde \gamma_{ij}}{3} \chi \tilde \gamma^{kl} \A_k \A_l \,, \\
\fl 4\pi \chi^{-1} j^i & = & \tilde \gamma^{li} \chi^{-3/2} \epsilon^F_{ljk} E^j B^k
+ \mu_{\rm V}^2 \Aphi \tilde \gamma^{ik} \A_{k} \,, \\
\fl 8\pi E & = &\chi^{-1} \tilde \gamma_{ij} \left( E^i E^j + B^i B^j \right) 
+ \mu_{\rm V}^2 \left(
\Aphi^2 + \chi \tilde \gamma^{kl} \A_k \A_l
\right) \,.
\end{eqnarray}

\bibliographystyle{iopart-num}
\bibliography{Proca_Bib}

\end{document}